\pdfoutput=1
\documentclass[journal]{IEEEtran}

\usepackage{amsmath,amsfonts}
\usepackage[ruled,vlined,linesnumbered]{algorithm2e}
\usepackage{array}
\usepackage{subcaption}
\usepackage{textcomp}
\usepackage{stfloats}
\usepackage{url}
\usepackage{verbatim}
\usepackage{graphicx}
\usepackage{cite}
\usepackage{multirow}
\usepackage{lipsum}
\usepackage{mathtools}
\usepackage{cuted}
\usepackage{breqn}
\usepackage[english]{babel}
\usepackage[autostyle]{csquotes}
\usepackage[acronym,toc,shortcuts]{glossaries}
\usepackage{comment}
\usepackage{siunitx}
\usepackage{textgreek}
\usepackage{xcolor}

\begin{document}

\title{Secure Cluster-Based Hierarchical Federated Learning in Vehicular Networks}

\author{M. Saeid HaghighiFard,~\IEEEmembership{Graduate Student Member,~IEEE} and Sinem Coleri,~\IEEEmembership{Fellow,~IEEE}
		
\thanks{M. Saeid Haghighifard and Sinem Coleri are with the Department of Electrical and Electronics Engineering, Koc University, Istanbul, Turkey, email: \{mhaghighifard21, scoleri\}@ku.edu.tr. This work is supported by the Scientific and Technological Research Council of Turkey Grant number 119C058 and Ford Otosan.}
}

\markboth{Journal of \LaTeX\ Class Files,~Vol.~14, No.~8, August~2021}%
{Shell \MakeLowercase{\textit{et al.}}: A Sample Article Using IEEEtran.cls for IEEE Journals}

 \maketitle
%%========== Abstract ==========%%
\IEEEtitleabstractindextext{%
    \begin{abstract}

Hierarchical Federated Learning (HFL) has recently emerged as a promising solution for intelligent decision-making in vehicular networks, helping to address challenges such as limited communication resources, high vehicle mobility, and data heterogeneity. However, HFL remains vulnerable to adversarial and unreliable vehicles, whose misleading updates can significantly compromise the integrity and convergence of the global model. To address these challenges, we propose a novel defense framework that integrates dynamic vehicle selection with robust anomaly detection within a cluster-based HFL architecture, specifically designed to counter Gaussian noise and gradient ascent attacks. The framework performs a comprehensive reliability assessment for each vehicle by evaluating historical accuracy, contribution frequency, and anomaly records. Anomaly detection combines Z-score and cosine similarity analyses on model updates to identify both statistical outliers and directional deviations in model updates. To further refine detection, an adaptive thresholding mechanism is incorporated into the cosine similarity metric, dynamically adjusting the threshold based on the historical accuracy of each vehicle to enforce stricter standards for consistently high-performing vehicles. In addition, a weighted gradient averaging mechanism is implemented, which assigns higher weights to gradient updates from more trustworthy vehicles. To defend against coordinated attacks, a cross-cluster consistency check is applied to identify collaborative attacks in which multiple compromised clusters coordinate misleading updates. Together, these mechanisms form a multi-level defense strategy to filter out malicious contributions effectively. Simulation results show that the proposed algorithm significantly reduces convergence time compared to benchmark methods across both 1‑hop and 3‑hop topologies. Under combined noise and gradient ascent attacks in the 1-hop scenario, the proposed algorithm reduces convergence time by 17.1\% compared to cosine‑similarity and Z‑score defenses, and by 8.7\% over the combined Z‑score and cosine strategy. In the 3‑hop setting under the same attacks, our proposed algorithm still leads with a 15.6\% reduction against Z‑score‑only methods and a 7.1\% improvement over the combined defense. These results demonstrate that the proposed framework consistently delivers superior resilience, accelerating convergence by up to 17\% even under intense adversarial conditions.

    \end{abstract}
}

\maketitle
\IEEEdisplaynontitleabstractindextext

%%========== Word Index ==========%%

%%========== Introduction ==========%%
\section{Introduction}

\IEEEPARstart{V}{ehicular} Ad hoc Networks (VANETs) are increasingly adopting machine learning (ML) algorithms to improve safety and efficiency in transportation systems. These algorithms use sensor data from LIDAR, RADAR, and vehicle-mounted cameras to enable intelligent decision-making. They support applications such as location-based services, traffic flow prediction and control, and autonomous driving. Traditionally, centralized ML methods train neural networks using large datasets collected from vehicle edge devices. In contrast, federated learning (FL) moves model training to the edge, reducing transmission overhead, and preserving user privacy. In FL, edge devices send parameter gradients to a central server instead of raw data. The server aggregates these gradients to update the global model, which is then redistributed to the devices for continued training. This process is repeated iteratively until the model converges~\cite{FLBook.ch2}.

%%========== HFL ==========%%

In recent years, Hierarchical Federated Learning (HFL) has emerged as a promising approach to address the challenges of deploying FL in vehicular networks. In HFL, a central cloud server coordinates global model training with the assistance of edge servers that act as intermediaries for model aggregation. This hierarchical approach helps mitigate issues related to limited communication resources, vehicular mobility, and data heterogeneity~\cite{9148862,9054634}. By enabling vehicles to transmit local model updates to nearby edge servers instead of directly to the cloud, HFL reduces communication overhead and expands coverage. It is also more effective in managing the dynamic nature of vehicular mobility and the non-independent and identically distributed (non-IID) data across different edge servers. Several studies have demonstrated the practical benefits of HFL:~\cite{chen2024mobility} demonstrates that HFL can leverage vehicular mobility to accelerate convergence in the presence of data heterogeneity, as higher speeds enable more diverse data exchange and faster data fusion across nodes;~\cite{chen2021semiasynchronous} proposes a semi-asynchronous HFL framework for Cooperative Intelligent Transportation Systems (C-ITS), improving communication efficiency in the presence of heterogeneous road vehicles;
and~\cite{haghighifard2024hierarchical} introduces a Cluster-based HFL (CbHFL) approach that uses multi-hop clustering and customized metrics to address both mobility and non-IID data, achieving significant improvements in accuracy and convergence time while maintaining an acceptable level of packet overhead. Despite these advances, security vulnerabilities, especially against adversarial attacks, remain largely unresolved. One of the most critical threats is model poisoning, where adversaries manipulate local model updates sent from vehicles to edge servers or the cloud. The decentralized nature of FL and absence of strict monitoring of individual client updates provide an ideal setting for adversaries to execute such attacks undetected. Since FL assumes that most client updates are benign, the aggregation process is highly vulnerable to slight adversarial modifications, which can accumulate over time and severely disrupt global model convergence~\cite{10024252,10274102,9857216}.

%%========== Anomaly detection ==========%%

Several defense strategies have been proposed to address security threats in FL. Among them, anomaly detection mechanisms have been developed to distinguish between malicious and benign model behavior in response to model poisoning attack~\cite{9751538, sun2021flwbc, 9734158, galanis2024defendingdatapoisoningattacks}.~\cite{9751538} introduces a differential privacy-exploited stealthy model poisoning (DeSMP) attack and proposes a reinforcement learning (RL)–based defense mechanism that dynamically adjusts privacy levels according to statistical metrics derived from client updates. Similarly,~\cite{sun2021flwbc} presents a client-based defense mechanism called White Blood Cell for FL (FL-WBC), which identifies vulnerable regions in the parameter space during local training and actively perturbs them to neutralize the effect of poisoning.~\cite{9734158} proposes filtering out anomalous models by assessing the similarity in historical update patterns. Complementing anomaly detection, dynamic client selection has been proposed as an additional layer of defense to enhance the robustness of FL systems. This approach optimizes the selection of participating vehicles in each training round based on the output of anomaly detection algorithms.~\cite{galanis2024defendingdatapoisoningattacks} proposes a method that monitors the update history of each client and statistically evaluates its performance using Z-Score. Clients whose updates fall below a predefined performance threshold are eliminated from the aggregation process. \cite{Khorramfar_2023} introduces a Credit-Based Client Selection (CBCS) method, ranking clients based on the accuracy of their past contributions. Clients with higher credit scores are preferred, while low-scoring clients are scrutinized or eliminated. In~\cite{yaldiz2023secure}, CosDefense uses cosine similarity to compare local updates with the global model, flagging and excluding clients whose updates deviate significantly, indicating potential malicious behavior. While eliminating persistently malicious participants through client elimination can reduce their adverse influence, it also presents challenges, particularly in non-IID settings. Due to the inherent heterogeneity of client data distributions, even benign clients may sometimes produce updates that are mistakenly classified as malicious, which reduces participation rates and compromises both performance and fairness. Moreover, existing defense approaches are not designed to exploit the hierarchical structure of FL in dynamic vehicular environments. Implementing dual-layer anomaly detection adds additional computational overhead and increases the likelihood of misclassification, especially given the mobility and variability of vehicular clients. Dynamic client selection in HFL remains underexplored, and its integration with anomaly detection has not been adequately investigated. 
%This gap highlights the critical need for novel methods that combine robust anomaly detection with adaptive client selection, specifically designed for hierarchical FL in highly dynamic vehicular contexts.

%%========== Novelty of the paper ==========%%

This paper proposes a novel framework that integrates dynamic client selection with statistical anomaly detection to effectively defend against model poisoning attacks in an HFL structure deployed over cluster-based VANETs.
The approach is designed to enhance both the robustness and reliability of the global model in adversarial and dynamic vehicular environments.
The key original contributions of this work are as follows:

\begin{itemize}
\item We propose a robust framework that defends against model poisoning attacks by integrating two statistical evaluation methods within an HFL structure for vehicular networks. Cosine similarity is used to measure the alignment between individual vehicle updates and the global model direction, while Z‑score analysis measures deviation in update norms from the collective average. This dual approach enables accurate differentiation between benign and malicious updates.

%ensuring that only trustworthy updates are aggregated into the global model.
\item We develop a novel dynamic client selection algorithm integrated with statistical anomaly detection to enhance the security of HFL in vehicular networks. Vehicles are clustered and assigned reliability scores based on three key metrics: historical accuracy, contribution frequency, and anomaly records. Historical accuracy prioritizes vehicles consistently demonstrating reliable cooperation during the FL process. The contribution frequency is used to identify vehicles with strong communication links and robust hardware. Meanwhile, anomaly records help pinpoint vehicles that exhibit minimal or no abnormal behavior. High-scoring vehicles are selected for participation, while those with suspicious behavior are temporarily excluded and later reassessed.
%This integrated mechanism not only preserves the integrity and accuracy of the global model but also fosters a secure, resilient, and collaborative learning environment in dynamic vehicular settings.

\item We propose an adaptive thresholding mechanism that continuously fine-tunes detection criteria based on the latest updated statistics. Instead of relying on fixed thresholds, our approach leverages real-time feedback from historical accuracy metrics to dynamically adjust the cosine similarity threshold. This adaptive strategy ensures that our dynamic client selection and filtering processes remain robust against unforeseen attack patterns.

\item We introduce a novel cross-cluster consistency check mechanism to further improve the integrity of the FL process in vehicular networks. By comparing aggregated updates across different clusters, the method effectively identifies anomalous clusters that may not be detected at the individual client level.

%, thereby significantly bolstering the overall robustness and reliability of the global model against adversarial perturbations.
\item We introduce a reliability-based weighted gradient averaging strategy to enhance model aggregation. Vehicles with higher reliability scores are given greater influence during global model updates, thereby reducing the impact of untrustworthy updates.
%By emphasizing reliable contributions, this method mitigates the impact of malicious or low-quality updates and helps maintain the integrity of the FL process.

\item We demonstrate through extensive simulations that our proposed framework significantly outperforms previously proposed cosine similarity and Z-score-based attacker detection algorithms in terms of convergence time and resilience under diverse attack scenarios.
\end{itemize}
%%========== Outline ==========%%

The subsequent sections of this paper are structured as follows. Section II provides the system model. Section III introduces the proposed secure HFL using anomaly detection and dynamic client selection algorithms. Section IV presents a comprehensive performance evaluation of the proposed algorithm compared to existing methods, across varying network sizes and transmission ranges. Finally, Section VI presents concluding remarks and future research directions.

%%========== System Model ==========%%
\section{System Model}
%%========== Cluster-Based Federating Leaning Network Model ==========%%
\subsection{Cluster-Based Hierarchical Federating Learning Network Model}

The network comprises a fleet of vehicles, each equipped with dual communication interfaces that support both vehicle-to-vehicle (V2V) and vehicle-to-infrastructure or vehicle-to-network (V2I/V2N) communications. These interfaces comply with standards such as IEEE 802.11p for Dedicated Short-Range Communication (DSRC)\cite{5888501}, IEEE 802.11bd for Wireless Access in Vehicular Environments (WAVE)\cite{9779322}, LTE-based Device-to-Device (D2D)\cite{7497762}, and 5G New Radio V2X (NR V2X)\cite{9392787}. 
%This robust communication infrastructure ensures efficient and reliable data exchange, enabling real-time information sharing essential for autonomous driving and intelligent transportation systems.

In this dynamic vehicular environment, vehicles form clusters using a clustering algorithm designed to balance data diversity and mobility. This ensures stable cluster head selection and robust model convergence, even under non-IID data distributions \cite{haghighifard2024hierarchical}. Within each cluster, vehicles take on one of two roles: Cluster Member (CM) or Cluster Head (CH). These roles adapt in real time to changes in network conditions. For example, if the performance of a CH is degraded due to high mobility or poor communication quality, a new CH is elected based on communication stability and computational capacity. This mechanism ensures that clusters remain adaptive and resilient to topology changes and vehicle dynamics.

The system employs hierarchical federated learning with a multi-tiered aggregation structure to optimize communication and training efficiency. Each CM acts as a client that generates model updates based on its local non-IID data. These updates, typically in the form of gradients, are transmitted to their respective CHs, which serve as intermediary aggregators. CHs collect and aggregate updates from the CMs within their cluster and forward the results to the Evolved Packet Core (EPC). The EPC acts as the central aggregation entity, responsible for performing global model aggregation across all clusters. The updated global model is then disseminated back to CHs and subsequently to CMs. This hierarchical structure significantly reduces the communication overhead at the EPC by limiting direct transmissions, while also improving the efficiency and speed of model convergence.

Each vehicle maintains a Vehicle Information Base (VIB), a dynamic data repository containing key operational parameters, including direction, location, velocity, cluster role, and local model updates. The VIB is updated upon operational changes or in response to events like receiving a ``$HELLO\_PACKET$" or a model update request. The ``$HELLO\_PACKET$" contains parameters such as vehicle direction, position, speed, current cluster affiliation, the connecting node to the cluster, cosine similarity of the FL model parameters, and relative velocity compared to surrounding vehicles. This information supports accurate clustering and ensures effective CH selection and model aggregation.

%%========== Attack Model ==========%%
\subsection{Attack Model}

We consider two types of model poisoning attacks: Gaussian noise and gradient ascent attacks. These attacks pose serious challenges to HFL in vehicular networks due to the highly dynamic nature of these environments. Vehicles frequently join and leave clusters, giving adversaries repeated opportunities to inject Gaussian noise, either through environmental disturbances or intentional manipulation, and to manipulate gradient updates to compromise system performance~\cite{shokri2015privacy}. The hierarchical structure of HFL, where CHs aggregate updates from CMs before relaying them to the EPC, adds an additional layer of vulnerability. This additional aggregation step provides adversaries with more opportunities to inject malicious updates. Moreover, the use of wireless communication channels exposes the system to eavesdropping and tampering, facilitating gradient ascent attacks that guide global models away from optimal solutions~\cite{li2019federated}.
 Furthermore, because vehicles generate non-IID data under diverse driving conditions, the effects of adversarial modifications are often unevenly distributed, with some clusters more severely impacted than others, compromising model integrity~\cite{bhagoji2019analyzing}. 
%%========== Gaussian Noise Attack ==========%%
\subsubsection{Gaussian Noise Attack}

In Gaussian noise attacks, adversarial vehicles add random noise to their local updates before sending them to the CH or EPC. This is a form of Byzantine attack, where the adversarial update  $g_{adv}$ from a malicious vehicle is given by
\begin{equation} \label{eq:1}
g_{\text{adv}} = g_{\text{benign}} + \mathcal{N}(m, \sigma^2 \mathbf{I}),\tag{1}
\end{equation}
where $g_{benign}$  is the genuine gradient update computed on the vehicle's local data, $\mathcal{N}(m, \sigma^2 \mathbf{I})$ represents Gaussian noise with mean $m$ and variance $\sigma^2$, and $I$ is the identity matrix. $\sigma^2$ controls the intensity of the noise injected.

By introducing Gaussian noise, adversarial vehicles aim to disrupt the learning process by making the aggregated gradients less representative of the true underlying data distribution. This random noise can cause the global model to converge slowly or even diverge, leading to degraded performance.
Gaussian attacks are dangerous because the added noise resembles natural variations in gradient updates, making it challenging to detect using simple anomaly detection methods that rely on statistical deviations. The randomness of the noise can mask the malicious intent, especially in networks with high variability among client updates.
%%========== Gradient Ascent Attack ==========%%

\subsubsection{Gradient Ascent Attack}
Gradient ascent attack is a more aggressive and targeted adversarial strategy where malicious vehicles deliberately manipulate model updates. Unlike benign gradient descent updates, which aim to minimize the loss function, gradient ascent attacks involve adversaries intentionally maximizing the loss function to steer global model parameters away from their optimal values~\cite{bhagoji2019analyzing}.

The malicious update $g_{adv}$ in a gradient ascent attack is designed to reverse the intended optimization direction. Let the benign update be given by
\begin{equation} \label{eq:2}
g_{\text{benign}} = -\eta \nabla L(\theta)\tag{2},
\end{equation}
where $\nabla L (\theta)$ is the gradient of the loss function with respect to model parameters $\theta$ and $\eta>0$ is the learning rate or scaling factor. To reverse this update, the adversarial client sends:
\begin{equation} \label{eq:3}
g_{\text{adv}} = g_{\text{benign}} + 2\eta \nabla L(\theta) = \eta \nabla L(\theta).\tag{3}
\end{equation}
This formulation clearly shows that the adversarial update is directed along $\nabla L(\theta)$, thereby maximizing the loss function.

The gradient ascent attack is particularly effective because it directly counteracts the beneficial updates from honest vehicles. Even a small proportion of adversarial vehicles can have a disproportionately large negative impact on the global model.
To counteract this, CHs need to detect and exclude these malicious vehicles from aggregation in HFL.
%%===== Secure HFL using Anomaly Detection and Dynamic Client Selection =====%%
\section{Secure HFL using Anomaly Detection and Dynamic Client Selection}

We propose a robust approach for dynamic client selection and anomaly detection within a cluster-based HFL framework for vehicular networks. This method is designed to defend against both Gaussian noise and gradient ascent attacks. To detect Gaussian noise attacks, we apply Z-score analysis on the norms of model updates. In typical scenarios, benign updates have norms that cluster around a mean value. When Gaussian noise is introduced, the update norms often deviate significantly from this mean, resulting in Z-scores that exceed a predefined threshold. Such updates are flagged as potentially malicious. To counter gradient ascent attacks, we use cosine similarity to assess the directional alignment between individual vehicle updates and the overall expected model direction. In a standard setting, benign updates are well aligned with the consensus update direction. In contrast, a low cosine similarity indicates that the update is misaligned, suggesting an adversarial attempt to intentionally steer the local or global model away from its intended direction by maximizing the loss function. By combining these two detection methods, we can identify both abnormal magnitudes and directional inconsistencies in updates, key signs of malicious behavior. To further improve detection accuracy, these statistical techniques are combined with vehicle reliability metrics, including historical accuracy, contribution frequency, and anomaly history. These reliability scores guide dynamic client selection, ensuring that only trustworthy and high-quality updates are included in the global model aggregation.
%%========== Detection Criteria ==========%%
\subsection{Detection Criteria}
%%========== Z-score ==========%%
\subsubsection{Z-score}
Z-score is used to identify outliers or anomalous behaviors by calculating how many standard deviations the gradient norm of a vehicle deviates from the mean of all vehicle gradient norms. Specifically, for each vehicle, the Z-score of its gradient norm is calculated to assess whether it significantly deviates from the collective average. This statistical approach allows us to detect vehicles whose updates might be corrupted due to faulty sensors, data errors, or malicious intent, such as in Gaussian attacks, where random noise is added to disrupt the learning process. By filtering out these extreme values, the Z-score helps maintain the quality of the aggregated model and prevents performance degradation~\cite{chandola2009anomaly}. 

For each CM $k$ at iteration $i$, the Z-score $Z\_Scores_{k,i}$ is calculated as:
\begin{equation} \label{eq:4}
Z\_Scores_{k,i} = \frac{Norms_{k,i} - Mean\_Norm_{i}}{Std\_Norm_{i}}\tag{4},
\end{equation}
where 
\[
Norms_{k,i} = \| g_{k,i} \|
\]
is the \( \ell_2 \)-norm of the gradient update \( g_{k,i} \) from vehicle \( k \) at iteration $i$. Here, $Mean\_Norm_{i}$ and $Std\_Norm_{i}$ denote the mean and standard deviation of the gradient norms across vehicles at iteration \( i \), respectively. These statistical measures capture the overall distribution of gradient norms, enabling the identification of outlier vehicles whose updates deviate significantly from the average. 
% \begin{equation} \label{eq:5}
% Mean\_Norm_{i} = \frac{1}{n} \sum_{k=1}^{n} Norms_{k,i}\tag{5}
% \end{equation}
% \begin{multline*} \label{eq:6}
% Std\_Norm_{k,i} = \\
% \sqrt{\frac{1}{n} \sum_{k=1}^{n} (Norms_{k,i} - Mean\_Norm_{i})^2}\tag{6}
% \end{multline*}
% which $n$ is the number of vehicles. 
%%========== at CH ==========%%

For each CH $k$ at iteration $i$, the Z-score $Z\_Scores_{k,i}$ is computed in a manner similar to that used for CM. However, the definition of the norm $Norm_{k,i}$ is adjusted to accommodate the CH's role in the aggregation process. Specifically, the norm for a CH is defined as the Euclidean distance between the CH's local aggregated update $\theta_{CH,i}$ and the global model $\theta_{global,i}$ as follows:
\begin{equation} \label{eq:7}
Norms_{CH,i} = \|\theta_{CH,i} - \theta_{global,i}\|\tag{5}.
\end{equation}
This formulation allows the Z-score to effectively measure deviations in the aggregated updates relative to the global model. By assessing this deviation, the system can identify abnormal aggregation behavior, which may result from compromised updates contributed by malicious CMs or malicious manipulation by the CH itself.
%%========== Cosine Similarity ==========%%
\subsubsection{Cosine Similarity}
Cosine similarity is used to assess the directional consistency of the gradient update of each vehicle relative to the mean gradient of all vehicles by measuring the cosine of the angle between them. A low cosine similarity indicates that the vehicle's update is inconsistent with the collective behavior, potentially signaling adversarial actions like gradient ascent attacks, where clients intentionally push the model parameters away from optimal values. By focusing on the direction rather than the magnitude of the gradients, cosine similarity helps detect such anomalies even when the gradient norms appear normal, adding an additional layer of defense against sophisticated attacks. 

For each CM \( k \) at iteration \( i \), the cosine similarity $Cos\_Sim_{k,i}$ is computed as:
\begin{equation} \label{eq:8}
Cos\_Sim_{k,i} = \dfrac{g_{k,i} \cdot Mean\_grad_{i}}{\|g_{k,i}\| \cdot \|Mean\_grad_{i}\|}\tag{6},
\end{equation}
where \( g_{k,i} \) is the gradient update from CM \( k \) at iteration \( i \), $Mean\_grad_i = \frac{1}{N} \sum_{j=1}^{N} g_{j,i}$  is the average gradient vector of all CMs within the cluster at iteration \( i \) and \( N \) is the number of valid CMs in a cluster after Z-score filtering. This calculation measures how closely each CM’s gradient aligns with the collective learning direction of the cluster.
%%========== at CH ==========%%

For CHs, cosine similarity evaluates the temporal consistency of the CH's aggregated model updates relative to the global model. By comparing the direction of consecutive aggregated updates, we detect anomalies at the aggregation level that may result from malicious activities or significant CH behavior deviations. Specifically, for each CH $j$ at iteration $i$, cosine similarity is calculated between the current aggregated update $\theta_{j,i}$ and the previous aggregated update $\theta_{j,i-1}$ relative to the previous iteration global model $\theta_{global,i-1}$ as follows:
\begin{multline*} \label{eq:9}
Cos\_Sim_{j,i} = \\
\dfrac{(\theta_{j,i} - \theta_{global,i-1}) \cdot (\theta_{j,i-1} - \theta_{global, i-1})}{\|\theta_{j,i} - \theta_{global,i-1}\| \cdot \|\theta_{j,i-1} - \theta_{global,i-1}\|}\tag{7}.
\end{multline*}
This cosine similarity measures the consistency of the CH’s updates over time, ensuring that the direction of model adjustments remains stable and aligned with the global learning objectives.
%%========== cross-cluster consistency check ==========%%

To further enhance security, the EPC performs a cross-cluster consistency check using cosine similarity to detect collaborative attacks, where multiple compromised CHs may coordinate to submit malicious or misleading updates. If several CHs submit similar updates that are inconsistent with the rest of the clusters, it could indicate a coordinated attack to skew the global model. In such cases, flagging these CHs ensures that the aggregated updates from CHs align with the overall learning objective and helps detect any anomalies or malicious activities at the cluster level. Specifically, after the EPC receives the aggregated model updates $\theta_{i}$ from each CH at iteration \( i \), it performs a consistency check by computing the cosine similarity between the updates of different clusters. The cross-cluster cosine similarity $Cos\_Sim_{p,q,i}$ between clusters \( p \) and \( q \) is calculated as:
\begin{multline*} \label{eq:10}
Cos\_Sim_{p,q,i} =\\
\frac{(\theta_{p,i} - \theta_{global,i-1}) \cdot (\theta_{q,i} - \theta_{global,i-1})}{\| \theta_{p,i} - \theta_{global,i-1} \| \times \| \theta_{q,i} - \theta_{global,i-1} \|}\tag{8},
\end{multline*}
where $\theta_{p,i}$ and $\theta_{q,i}$ are the model updates from CHs of clusters \( p \) and \( q \) at iteration \( i \), respectively, and $\theta_{global,i-1}$ is the global model from the previous iteration.

To quantify the overall consistency of a cluster's update with the rest of the network, we calculate the average cross-cluster cosine similarity $\overline{Cos\_Sim}_{p,i}$ for cluster \( p \) as:
\begin{equation} \label{eq:11}
\overline{Cos\_Sim}_{p,i} = \frac{1}{C - 1} \sum_{\substack{q=1 \\ q \neq p}}^{C} Cos\_Sim_{p,q,i}\tag{9},
\end{equation}
where \( C \) is the total number of clusters. By computing this average, we assess how well the update of the cluster $p$ aligns with the updates of the other clusters in the same iteration. This mechanism ensures that the clusters that contribute to the global model align with the overall learning objective, improving the robustness and reliability of the FL.
%%========== Reliability Metrics ==========%%
\subsubsection{Reliability Metrics}
We integrate the outputs of Z-score and cosine similarity-based anomaly detection into a set of reliability metrics to evaluate and prioritize vehicles based on their historical performance. The first metric, Historical Accuracy, assesses the long-term performance of each vehicle by measuring the average accuracy of its local model over multiple communication rounds using a validation dataset. The second metric, Contribution Frequency, measures how consistently a vehicle's updates have been accepted into the aggregation process based on Z-score and cosine similarity assessments, while also accounting for packet losses due to channel quality, which may prevent a vehicle from contributing at various learning stages and thereby lower its overall frequency of contribution. The third metric, Anomaly Record, tracks the number of times a vehicle's updates have been flagged as anomalous due to significant deviations detected by Z-score or misalignment identified by cosine similarity. By integrating these reliability metrics with the statistical evaluations from Z-score and cosine similarity, our approach enhances the robustness of anomaly detection and improves the overall reliability of the federated learning process.
%%========== Historical accuracy ==========%%

Historical accuracy reflects the average performance of the local model of a vehicle in multiple communication rounds. This metric is calculated by the EPC and all CHs using the accuracy of the model on a validation dataset during each round~\cite{Goodfellow2016}. Accuracy is defined as the percentage of correct predictions made by the model when tested on unseen validation samples. Higher historical accuracy indicates that the vehicle consistently contributes high-quality updates to the global model. The historical accuracy of vehicle $k$ at round $i$ is computed as:
\begin{multline*} \label{eq:12}
    Historical\_Accuracy_{{k,i}} = Total\_Accuracy_{k,i}/i = \\
    (1/i) * \sum_{n=1}^{i}(Accuracy\_of\_Contribution_{k,n}),\tag{10}
\end{multline*}
where $i$ represents the $i$th communication round considered as a single FL iteration within the FL procedure,
$Total\_Accuracy_{k,i}$ represents the sum of all $Accuracy\_of\_Contribution_{k,n}$ values across all communication rounds up to the current round $i$, and $Accuracy\_of\_Contribution_{k,n}$ is the accuracy of vehicle $k$ at round $n$, evaluated by the CH or EPC as:
\begin{multline*}\label{eq:13}
Accuracy\_of\_Contribution_{k,n} = \\
\left( \frac{\sum_{j=1}^{N_{\text{val}}} \delta_j}{N_{\text{val}}} \right) \times 100\%.\tag{11}
\end{multline*}
where $D_{\text{val}} = \{(x_j, y_j)\}_{j=1}^{N_{\text{val}}}$ is the validation dataset, $x_j$ is the input data, $y_j$ is the true label for the $j$-th sample, $N_{\text{val}}$ is the number of samples in the validation dataset, 
\begin{equation}\label{eq:15}
\delta_j = \mathbb{I}[\hat{y}_j = y_j] = 
\begin{cases} 
1, & \text{if } \hat{y}_j = y_j, \\
0, & \text{if } \hat{y}_j \neq y_j \tag{12},
\end{cases}
\end{equation}
$\hat{y}_j = \theta_{k,n}(x_j)$ is the predicted label for the $j$-th sample by using 
the vehicle's updated model $\theta_{k,n}$.
By calculating the number of correct predictions, we obtain an accurate measure of the vehicle's model performance on unseen data.

%%========== Contribution frequency ==========%%

Contribution frequency measures how regularly a vehicle participates in FL rounds. A higher frequency implies that the vehicle is consistently active and dependable. This metric, however, is affected not only by the vehicle’s willingness to participate but also by its channel quality. Poor channel conditions may lead to failed or delayed transmissions, reducing the effective contribution frequency. In contrast, vehicles with stable and strong communication links are more likely to contribute updates consistently, reinforcing their reliability as participants in the FL process. The contribution frequency of vehicle $k$ is calculated by
\begin{equation}\label{eq:16}
    Contribution\_Freq_{k,i} = {Total\_Contributions_{k,i}}/{i},\tag{13}
\end{equation}
where $Total\_Contributions_{k,i}$ represents the number of contributions made by vehicle $k$ across all communication rounds up to the current round $i$ in the FL process.
%%========== Anomaly record ==========%%

Anomaly record tracks the frequency of a vehicle's contributions being flagged as anomalous through Z-score and cosine similarity analyses.  A high anomaly record suggests that the vehicle may be compromised or unreliable, and thus should be assigned a lower reliability score to protect the integrity of the learning process. The anomaly record of vehicle $k$ at round $i$ is calculated by
\begin{multline*} \label{eq:17}
    Anomaly\_Record_{k,i} = \\
    {Total\_Anomalous\_Contributions_{k,i}}/{i},\tag{14}
\end{multline*}
where $Total\_Anomalous\_Contributions_{k,i}$ refers to the total number of times the contributions of vehicle $k$ have been flagged as anomalous across all communication rounds up to the current round $i$.
%%========== Reliability score ==========%%

The final reliability score for each vehicle is determined by a weighted sum of the three metrics as given below:
\begin{multline*} \label{eq:18}
Reliability\_Score_{k,i} = \\
({Accuracy\_Weight \times Historical\_Accuracy_{k,i}})+\\
({Frequency\_Weight \times Contribution\_Freq_{k,i}})-\\
({Anomaly\_Weight \times Anomaly\_Record_{k,i}})\tag{15}
\end{multline*}
Assigning specific weights to these metrics allows for a flexible and context-sensitive evaluation of each vehicle's contributions in FL. This weighting system can be customized to emphasize particular aspects of reliability based on the specific needs and goals of the FL application. As the network evolves, the reliability score is dynamically updated, ensuring that the selection process remains adaptive to changes in vehicle behavior and network conditions. Vehicles with higher reliability scores are prioritized for participation in model training rounds, thereby enhancing the overall robustness and accuracy of the global model. Vehicles identified as anomalous through Z-score and cosine similarity checks have their anomaly records updated, affecting their reliability scores. This feedback mechanism ensures that the system remains responsive to changes in vehicle behavior, reducing the impact of unreliable or malicious vehicles while promoting contributions from those who consistently provide valuable data.
%%========== Adaptive thresholding ==========%%
\subsubsection{Adaptive Thresholding Mechanism}

We introduce an adaptive thresholding mechanism for the cosine similarity metric to enhance the robustness of our anomaly detection process. This mechanism dynamically adjusts the cosine similarity threshold based on each vehicle’s historical accuracy,
ensuring that vehicles with consistently high performance are
held to stricter standards. High-performing CMs might become
less careful over time, leading to a gradual decline in update
quality. Without adaptive thresholding, these vehicles may
continue to pass the anomaly detection checks despite a drop
in performance, as the static thresholds do not adjust to their
historical contribution levels. The adaptive cosine similarity threshold $Cosine\_Sim\_Threshold_{adaptive,k}$ for vehicle \( k \) is defined as:
\begin{multline*} \label{eq:19}
Cosine\_Sim\_Threshold_{adaptive,k} =\\
Cosine\_Sim\_Threshold_{adaptive,k} - \delta,\tag{16}
\end{multline*}
where $\delta$ is a small positive adjustment parameter controlling the degree of threshold tightening. When a vehicle's $Historical\_Accuracy_{k,i}$ exceeds $High\_Threshold\_Up$ and the cosine similarity threshold is above the floor threshold $High\_Threshold\_Down$, the threshold is reduced by $\delta$. This adjustment continues until the threshold reaches $High\_Threshold\_Down$, at which point further reduction stops. This adjustment imposes stricter anomaly detection criteria on high-performing vehicles, ensuring that even minor deviations from their typical update patterns are detected.

%%========== Weighted Gradient Averaging ==========%%
\subsubsection{Weighted Gradient Averaging}
We implement a weighted gradient averaging mechanism to enhance the robustness and effectiveness of the model aggregation process in FL. This method assigns weights to each CM's gradient updates based on their computed reliability scores, ensuring that more reliable participants have a greater influence on the global model. Specifically, after the anomaly detection and reliability score adjustment steps, the aggregated gradient \( G_i \) is computed as a weighted average of the accepted gradients:
\begin{equation} \label{eq:20}
G_{i} = \frac{\sum_{k=1}^{N} Reliability\_Score_{k,i} \times g_{k,i}}{\sum_{k=1}^{N} Reliability\_Score_{k,i}},\tag{17}
\end{equation}
where $N$ represents the number of CMs in the cluster under that CH, $Reliability\_Score_{k,i}$ and $g_{k,i}$ denote the reliability score and gradient of the $k$th CM during the $i$ communication round, respectively.
It then updates its model parameters using a gradient descent step:
\begin{equation} \label{eq:23}
\theta_{CH,i} = \theta_{CH,i-1} - \eta\,G_{i},\tag{18}
\end{equation}
where $\theta_{CH,i-1}$ represents the CH’s model parameters carried over from the previous round.
This approach effectively amplifies the positive impact of trustworthy CMs while diminishing the influence of those with lower reliability. The normalization by the sum of reliability scores ensures that the aggregated gradient maintains an appropriate scale.

At the cluster level, a similar weighted averaging is performed when aggregating the updates from CHs at the EPC. The global model parameters \( \theta_{global,i} \) at iteration \( i \) are updated using the weighted average of the CHs' model parameters \( \theta_{CH,i} \):
\begin{equation} \label{eq:21}
\theta_{global,i} = \frac{\sum_{CH=1}^{C} Reliability\_Score_{CH,i} \times \theta_{CH,i}}{\sum_{p=1}^{C} Reliability\_Score_{CH,i}}\tag{19},
\end{equation}
where $Reliability\_Score_{CH,i}$ is the reliability score of CH after any adjustments due to cross-cluster consistency checks or anomaly penalty factors, $\theta_{CH,i}$ is the local update from CH at iteration \( i \) and $C$ is the number of clusters. This weighted averaging process enhances the overall robustness and effectiveness of the FL by prioritizing contributions from trustworthy clusters.
%%===== Anomaly Detection and Dynamic Vehicle Selection Algorithm in CHs and EPC =====%%
\subsection{Dynamic Anomaly Detection and Reliability-based Client Selection (DARCS) Algorithm}

In the proposed HFL framework for vehicular networks, the CHs and the EPC have distinct responsibilities, each supported by dedicated algorithms. CHs serve as intermediaries between CMs and the EPC. They manage communication within their clusters, perform local model aggregations, and execute anomaly detection mechanisms tailored to their respective clusters. The CH algorithm focuses on dynamic client selection and anomaly detection at the CM level, using individual vehicle reliability scores. 
On the other hand, the EPC serves as the central coordinator of the entire network. It aggregates model updates from all CHs to refine the global model and performs higher-level anomaly detection, including cross-cluster consistency checks and reliability evaluations of CHs based on aggregated metrics. Unlike CHs, the EPC algorithm operates at the cluster level and processes aggregated rather than individual vehicle updates. By employing different algorithms proposed for the specific functions and data accessibility of the CHs and EPC, this hierarchical approach effectively mitigates risks posed by malicious actors and unreliable data sources at different levels of the network hierarchy.
%%========== Algorithm 1 ==========%%
\begin{algorithm}[!]
\caption{Dynamic Anomaly Detection and Reliability-based Client Selection (DARCS) Algorithm at the CH}

\ForEach {$CM \in VIB$}{
    Initialize $Reliability\_Score_{CM,i}$ based on VIB;\\
    \If{$Block\_Flag_{CM} =$ TRUE \textbf{and} $block\_duration_{CM} < Unblock\_Time$}{
        Increment $block\_duration_{CM}$;\\
        \textbf{Continue} to next $CM$;\\
    }
    \Else{
        Reset $Block\_Flag_{CM}$ to FALSE;\\
    }
}
    
    Sort all $CM$ with $Block\_Flag_{CM} = FALSE$ by $Reliability\_Score_{CM,i}$ in descending order;\\
    Select top $SELECTED\_CLIENT\_PERCENTAGE$ of vehicles as $Selected\_Clients$;\\
    
\ForEach {$CM \in Selected\_Clients$}{
    Compute $Z\_Scores_{CM,i}$ by using~\eqref{eq:4};\\

    \If{$|Z\_Score_{CM,i}| < Z\_Score\_Threshold$}{
        Compute $Cos\_Sim_{CM,i}$ by using~\eqref{eq:8};\\
    }
    \Else{
        Set $Block\_Flag_{CM} =$ TRUE and $block\_duration_{CM} = 0$;\\
        Increment $Total\_Anomalous\_Contributions_{CM,i}$;\\
        \textbf{Continue} to next $CM$;\\
    }

    \If{$|Cos\_Sim_{CM,i-1}-Cos\_Sim_{CM,i}| > Cosine\_Sim\_Threshold_{adaptive,CM}$}{
        Reset CM parameters using $Memory_{CM,i-1}$
    }
    \Else{
        Increment $Total\_Contributions_{CM,i}$;\\
        Calculate $Accuracy\_of\_Contribution_{CM,i}$;\\
        Compute $Historical\_Accuracy_{CM,i}$ by using~\eqref{eq:12}, $Contribution\_Freq_{CM,i}$ by using~\eqref{eq:16}, $Anomaly\_Record_{CM,i}$ by using~\eqref{eq:17};\\
        Update $Reliability\_Score_{CM,i}$ by using~\eqref{eq:18};\\
        Record updated metrics in VIB;\\
    }
    \If{$Historical\_Accuracy_{CM,i} \geq High\_Threshold\_Up$ and $Cosine\_Sim\_Threshold_{adaptive, CM}$ $>$ $High\_Threshold\_Down$}{
        Adjust $Cosine\_Sim\_Threshold_{adaptive,CM}$ by using~\eqref{eq:19};\\
    }
}

Compute model update $\theta_{CH,i}$ by using~\eqref{eq:20} and ~\eqref{eq:23};\\
\ForEach{$CM\in VIB$}{
    Record all parameters in $Memory_{CM,i}$;\\
}
Send updated model parameters $\theta_{CH,i}$ to all CMs and EPC\;

\end{algorithm}
%%========== Algorithm 1 Explanation ==========%%

Algorithm 1 operates within the Cluster-based HFL framework and incorporates mechanisms to enhance robustness against unreliable or adversarial gradient updates from CMs at the CH. The primary challenge is efficiently aggregating these local updates while ensuring malicious contributions do not compromise the overall model convergence. The CH initializes the reliability scores for each CM based on the data in the VIB (Lines 1-2). At the beginning of each communication round, the algorithm assesses the participation eligibility of each CM. If a CM's $Block\_Flag_{CM}$ is set to $TRUE$ and the $block\_duration$ is less than the predefined $Unblock\_Time$, it indicates that the CM has been previously identified as malicious and is still within its blocking period. In this case, the $Unblock\_Time$ is incremented, and the CH skips further processing for this CM, proceeding to the next. If the condition is not met—meaning the CM is either not blocked, or the blocking duration has expired—the algorithm resets the $Block\_Flag_{CM}$ to $FALSE$, allowing the CM to participate in the current round, and the CH receives the update $g_{CM,i}$ from the CM (Lines 3-7).
In the subsequent stage, all CMs that have passed the blocking check are considered eligible. These eligible CMs are then sorted in descending order based on their reliability scores, and the top $SELECTED\_CLIENT\_PERCENTAGE$ are designated as $Selected\_Clients$. This selection process prioritizes the most reliable CMs for participation in the current training round, enhancing the overall model quality (Lines 8-9).

Next, Z-scores are calculated for the gradient norms of each CM, indicating how far a particular CM's gradient is from the mean in terms of standard deviations (Line 11). If the absolute value of $Z\_Scores_{CM,i}$ is less than the $Z\_Score\_Threshold$, the algorithm computes the cosine similarity between each valid CM gradient and the mean gradient to further ensure robustness (lines 12-13).
In contrast, if the absolute value of $Z\_Score_{CM,i}$ exceeds $Z\_Score\_Threshold$, it indicates an outlier or anomalous behavior. In this case, the algorithm sets the $Block\_Flag_{CM}$ to TRUE, initializes the block duration timer to zero, increments the $Total\_Anomalous\_Contributions_{CM,i}$ counter, and proceeds to skip this CM, moving on to the next CM in the cluster (Lines 14-17).

If the absolute difference between $Cos\_Sim_{CM}$ in the current and previous communication rounds exceeds $Cosine\_Sim\_Threshold_{adaptive,i}$, it indicates a substantial change that may serve as a potential sign of malicious intent, where a CM introduces significant deviations to disrupt the learning process. Upon detecting this anomaly, the CM's parameters are reset using $Memory_{CM,i-1}$, which contains the stored model parameters from the previous round. A memory mechanism replicates the most recent CM structure, ensuring that previous CM parameters remain accessible within the system. This approach prevents abrupt omissions of a CM based solely on cosine similarity differences, which could otherwise disrupt the learning process (Lines 18-19).

After filtering out unreliable CMs based on cosine similarity, the algorithm updates the contribution records for the remaining valid CMs. It calculates the $Accuracy\_of\_Contribution_{CM,i}$, updates the $Total\_Accuracy_{CM,i}$, and computes the $Historical\_Accuracy_{CM,i}$. It also computes the $Contribution\_Freq_{CM,i}$ and $Anomaly\_Record_{CM,i}$ to assess each CM's reliability over time. The $Reliability\_Score_{CM,i}$ is updated using these metrics, incorporating weights for accuracy, frequency, and anomaly records. This score reflects the CM's trustworthiness and influences its future participation and impact on the model aggregation (Lines 20-24). Subsequently, these updates are added to the VIB for further decision-making and calculation (Line 25).

The next step applies adaptive thresholding by adjusting the cosine similarity threshold for the high-performing CMs. This mechanism dynamically adjusts the cosine similarity threshold based on each CM's historical accuracy, ensuring that CMs with consistently high performance are held to stricter standards. Specifically, if a CM’s historical accuracy surpasses a predefined $High\_Threshold\_Up$, it indicates that the CM has consistently contributed reliable updates over time. The algorithm tightens the cosine similarity threshold for these high-performing CMs using~\eqref{eq:19} until it is not less than a predefined parameter $High\_Threshold\_Down$ to ensure sustained high-performance (Lines 26-27). The CH aggregates the accepted updates from CMs using weighted averaging based on reliability scores (Line 28). The current parameters of each CM are then stored in a memory buffer denoted as $Memory_{CM,i}$. Finally, the CH sends the average aggregated update $\theta_{CH,i}$ to the EPC and receives the global model update in return. It distributes the updated model parameters to all CMs in the cluster, completing the training round (Lines 29-31).
%%========== Algorithm 2 ==========%%
\begin{algorithm}[!]
\caption{Dynamic Anomaly Detection and Reliability-based Client Selection (DARCS) Algorithm at the EPC} 
    \ForEach {$CH \in VIB$}{
        \If{$Block\_Flag_{CH} =$ TRUE \textbf{and} $block\_duration_{CH} < Unblock\_Time$}{
            Increment $block\_duration_{CH}$\;
            \textbf{Continue} to next CH\;
        }\Else{
            Reset $Block\_Flag_{CH}$ to FALSE\;
        }
        Compute $Norms_{CH,i}$ by using~\eqref{eq:7};\\
        Compute $Z\_Score_{CH,i}$ by using~\eqref{eq:4}\;
        \If{$|Z\_Score_{CH,i}| < Z\_Score\_Threshold$}{
            Compute $Cos\_Sim_{CH,i}$ by using~\eqref{eq:9};
        }\Else{
            Set $Block\_Flag_{CH}=$ TRUE and $block\_duration_{CH} = 0$\;
            Increment $Total\_Anomalous\_Contributions_{CH,i}$;\\
            \textbf{Continue} to next CH\;
        }
        
         \If{$|Cos\_Sim_{CH,i-1}-Cos\_Sim_{CH,i}| > Cosine\_Sim\_Threshold_{adaptive,CH}$}{
        Reset CH parameters using $Memory_{CH,i-1}$
        }
         \Else{
         \ForEach{pair of CHs $(p, q)$ where $p \ne q$}{
            Compute $Cos\_Sim_{p,q,i}$ by using~\eqref{eq:10};
            } 
        \ForEach{$p$}{
        Compute $\overline{Cos\_Sim}_{p,i}$ by using~\eqref{eq:11};\\
        \If{$\overline{Cos\_Sim}_{p,i} < Cosine\_Sim\_Threshold_{cross}$}{
        Set $Block\_Flag_{CH} =$ TRUE and $block\_duration_{CH} = 0$\;
        Increment $Total\_Anomalous\_Contributions_{CH,i}$;\\
        \textbf{Continue} to next CH\;
        }
        }  
        }
        \Else{
            Increment $Total\_Contributions_{CH,i}$;\\
            Calculate $Accuracy\_of\_Contribution_{CH,i}$;\\
            Compute $Historical\_Accuracy_{CH,i}$ by using~\eqref{eq:12}, $Contribution\_Freq_{CH,i}$ by using~\eqref{eq:16}, $Anomaly\_Record_{CH,i}$ by using~\eqref{eq:17};\\
            Update $Reliability\_Score_{CH,i}$ by using~\eqref{eq:18};\\
            Record updated metrics in VIB;\\
        }
        
        \If{$Historical\_Accuracy_{CH,i} \geq High\_Threshold\_Up$ and $Cosine\_Sim\_Threshold_{adaptive, CH}$ $>$ $High\_Threshold\_Down$}{
        Adjust $Cosine\_Sim\_Threshold_{adaptive,CH}$ by using~\eqref{eq:19};\\
        }        
}
    \textbf{Aggregate} the accepted model parameters from CHs\;
    Compute global model update $\theta_{global,i}$ by using~\eqref{eq:21};\\
    \ForEach {$CH \in VIB$}{
    Record all parameters in $Memory_{CH,i}$;\\}
    Send updated global model parameters $\theta_{global,i}$ to all CHs\;
\end{algorithm}

%%========== Algorithm 2 Explanation ==========%%
Algorithm 2 operates within the same framework and improves robustness against unreliable or adversarial model updates from CHs at the EPC. The primary challenge is to efficiently aggregate these cluster-level updates while ensuring that malicious contributions do not compromise the convergence of the global model. Initially, the algorithm iterates over each CH in the VIB and manages blocked CHs the same way as Algorithm 1 (Lines 1–6). It then calculates $Norms_{CH,i}$, to quantify the deviation of each CH from the global model, updating the mean and standard deviation of these norms to build a statistical profile of the updates of the CHs. Using this profile, the algorithm computes Z scores for each CH model update (Lines 7–8) to identify outliers and track anomaly records (Line 9). For CHs passing the Z-score check, the cosine similarity between each CH’s update and the global model parameters is computed (Line 10). If a CH shows a significant deviation in cosine similarity between consecutive rounds, the algorithm reverts the parameters of that CH to the most recent trusted state, mirroring Algorithm 1 but with a specific index for CH vehicles (Lines 15-16).

The algorithm then performs a cross-cluster consistency check to ensure consistent aggregated updates from different CHs. For each pair of CHs, the cross-cluster cosine similarity is computed to measure the directional alignment between their updates. For each CH, it calculates the average cross-cluster cosine similarity with all other CHs. If a CH's average cross-cluster cosine similarity is below the predefined threshold $Cosine\_Sim\_Threshold_{cross}$, it indicates that the CH's update is inconsistent with those from other clusters, possibly due to anomalies or adversarial behavior. In such cases, the algorithm sets the $Block\_Flag_{CH}$ to $TRUE$, resets the $block\_duration_{CH}$ to zero, increments the $Total\_Anomalous\_Contributions_{CH,i}$ counter, and skips processing this CH (Lines 17-25). After filtering out unreliable CHs via these Z-score and cosine similarity evaluations, the algorithm updates the contribution records of each remaining CH following the same procedure as Algorithm 1 (Lines 26–31). Next, adaptive thresholding is applied by adjusting the cosine similarity threshold for high-performing CHs. If a CH's $Historical\_Accuracy_{CH,i}$ surpasses a predefined $High\_Threshold\_Up$, it indicates consistently high performance (Lines 32-33).

The EPC then aggregates the accepted model parameters from the remaining CHs to update the global model. The aggregation is performed using weighted averaging based on the CHs' reliability scores, ensuring that more reliable CHs have a more significant influence on the global model update. This weighted aggregation promotes the contributions of trustworthy CHs while mitigating the impact of less reliable ones, enhancing the overall performance and convergence of the global model (Lines 34-35). The subsequent step focuses on saving the current parameters of each CH into a designated memory buffer, referred to as $Memory_{CH,i}$ (Lines 36-37).

Finally, the EPC sends the updated global model parameters back to all CHs for distribution to their respective CMs, completing the training round. This dissemination ensures all participants have the latest model parameters, enabling them to continue the local training and updating process in subsequent rounds (Line 38).
%%========== Performance Evaluation ==========%%
\section{Performance Evaluation}

The simulations evaluate the effectiveness of the proposed Dynamic Anomaly Detection and Reliability-based Client Selection (DARCS) algorithm when integrated with the Cluster-based HFL (CbHFL) framework presented in~\cite{haghighifard2024hierarchical}. Our approach builds on CbHFL’s dynamic clustering mechanism, which adapts to highly mobile vehicular environments and enables multi-hop transmission of local updates to CHs. We compare the performance of the DARCS-enhanced CbHFL framework against four benchmark approaches. The first benchmark is the original CbHFL framework without any attacks. In this benchmark, vehicles are grouped into CMs and CHs, while the EPC acts as the central server for global model aggregation. This baseline serves as a reference for an effective HFL environment in a non-IID, highly dynamic vehicular setting. The second benchmark is a cosine similarity-based attacker detection algorithm (CosDefense)~\cite{yaldiz2023secure}, adapted for CbHFL. It identifies malicious vehicles by computing cosine similarity scores between the global model’s last-layer weights and local updates, excluding vehicles with significant deviations from the average. The third benchmark is Z-Score Detection~\cite{galanis2024defendingdatapoisoningattacks}, also adapted for CbHFL, which detects outliers based on Z-score analysis and removes updates that exceed a predefined threshold. The fourth benchmark combines cosine similarity and Z-score methods but does not incorporate reliability scores, adaptive thresholding, or cross-cluster consistency checks. This variant helps to isolate and evaluate the impact of additional mechanisms proposed in DARCS.

%The proposed DARCS algorithm outperforms these methods by integrating all relevant metrics within the CbHFL framework, offering robust anomaly detection and enhanced security against malicious attacks. By also comparing these approaches to a baseline scenario without attacks, the evaluation demonstrates DARCS’s superior ability to maintain performance and improve defense mechanisms in the presence of adversarial activities.
%%========== Simulation Setup ==========%%
\subsection{Simulation Setup}
The simulations utilize Python, with vehicle mobility generated by Simulation of Urban Mobility (SUMO) and data streaming via KAFKA. This setup creates a realistic vehicular network scenario for FL. SUMO models individual driver behaviors and traffic dynamics, while KAFKA is utilized to transmit packets and model parameters, enabling communication among vehicles. FL models are built using PyTorch. 
%%========== Simulation environment ==========%%

The simulation environment replicates a one km² area with two-lane roads, where vehicle movements are modeled using a Poisson process. Vehicles have speeds ranging from 10 m/s to 35 m/s, capturing a variety of mobility patterns. IEEE 802.11p is employed for vehicle-to-vehicle communication, while 5G NR is used for vehicle-to-infrastructure communication. For vehicle-to-vehicle links, the Winner+ B1 propagation model is applied as described in~\cite{9599363}. For vehicle-to-infrastructure communication, the Friis propagation model estimates signal attenuation and propagation characteristics between vehicles and 5G NR base stations, following~\cite{3gpptr38901}. FL training uses the MNIST dataset with stochastic gradient descent (SGD), under a non-IID data distribution across vehicles. The main evaluation metrics are model accuracy, indicating the predictive performance of the model~\cite{Goodfellow2016}, and convergence time, representing the number of communication rounds required to achieve model stability. Convergence is defined using a threshold $\epsilon$, where learning is considered complete if the improvement in model performance falls below $\epsilon$ over three consecutive rounds~\cite{mcmahan2017communication}.

%%========== Simulation Parameters ==========%%

A comprehensive set of simulations is conducted to evaluate the contribution of each component of the proposed algorithm under three scenarios: (i) 25 vehicles with a 100-meter transmission range, (ii) 25 vehicles with a 500-meter transmission range, and (iii) 50 vehicles with a 100-meter transmission range. In each case, 20\% of the vehicles are designated as attackers, employing harsh fake additive Gaussian noise (mean = 2, variance = 0.3), gradient ascent attacks, or a combination of both. The evaluation models a worst-case scenario where malicious vehicles begin attacking from the first communication round and continue without interruption. Both a single-hop configuration, allowing direct CM–CH communication, and a 3-hop setup, involving data relaying through intermediate CMs, are considered to assess performance under different network topologies.
%%========== Key Parameter Settings ==========%%
Key parameter settings for this scenario include a vehicle selection ratio of 75\% per training round, an unblock time fixed at 5 rounds, and specific thresholds: a $Z\_Score\_Threshold$ of 3, an $Cosine\_Sim\_Threshold_{adaptive,k}$ of 0.90, $High\_Threshold\_Up$ of 95\%, $High\_Threshold\_Down$ of 0.2, an Adjustment Factor ($\delta$) of 0.05, and a $Cosine\_Sim\_Threshold_{cross}$ of 0.9. The weighting of each parameter towards the reliability score is equal.

%To benchmark the proposed algorithm, three distinct aforementioned scenarios are analyzed under both single-hop and three-hop communication configurations. This setup tests the algorithm's robustness against adversarial tactics, specifically observing the effects of continuous network attacks on model integrity and performance. These evaluations provide insights into the resilience of FL models under varying adversarial conditions and network complexities.
%%========== Performance Evaluation of Proposed Algorithm ==========%%
\subsection{Performance Evaluation of Proposed Algorithm}
%%==========================FIG I=============================%%
\begin{figure}[!t]
     \centering
     \begin{subfigure}{0.55\textwidth}
         \centering
         \includegraphics[width=\textwidth]{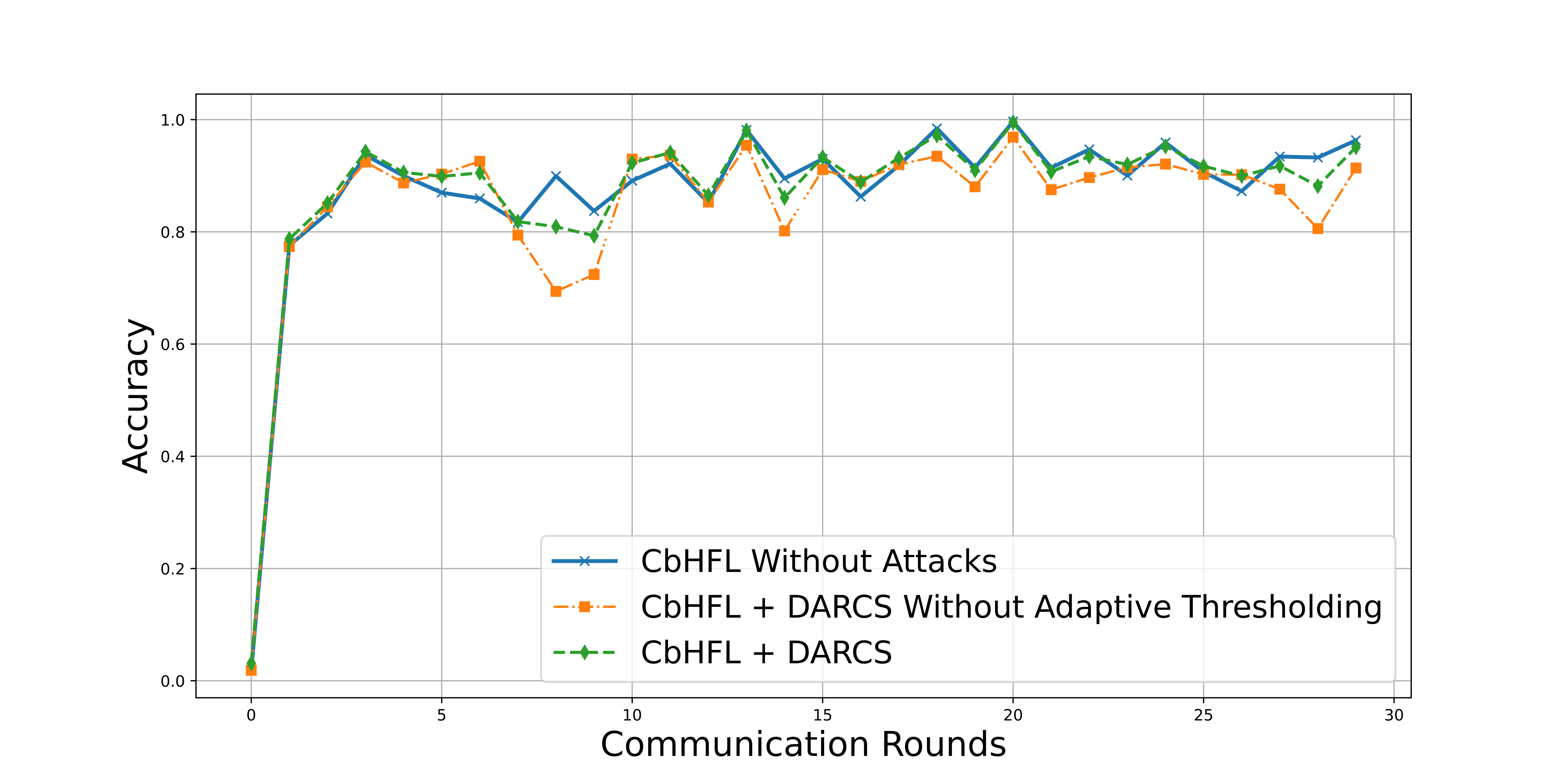}
         \subcaption*{(a)}
         \label{fig: Cont-0-0.3}
     \end{subfigure}
     \vfill
     \begin{subfigure}{0.55\textwidth}
         \centering
         \includegraphics[width=\textwidth]{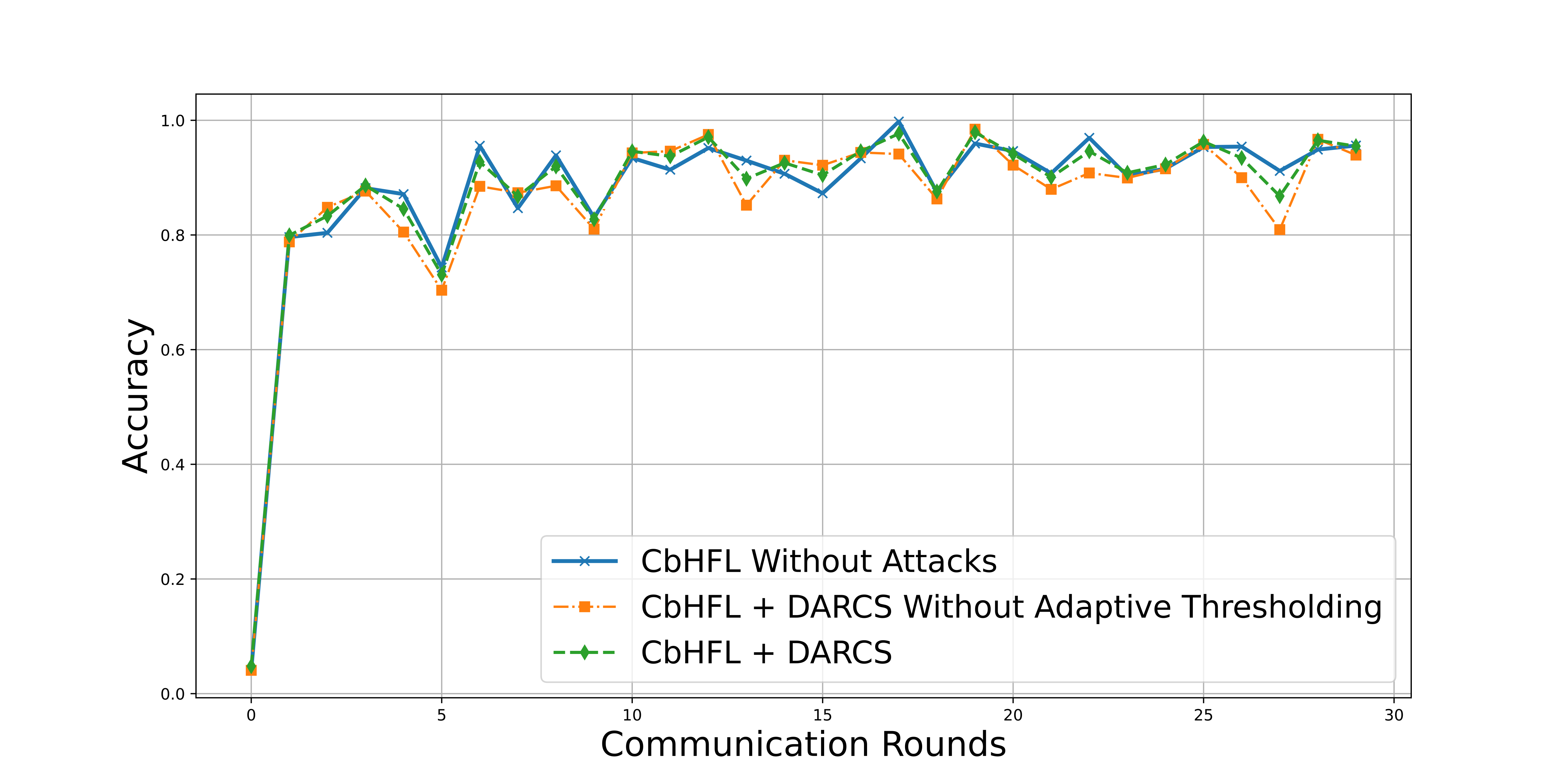}
         \subcaption*{(b)}
         \label{fig: Cont-Gradient}
     \end{subfigure}
     \vfill
     \begin{subfigure}{0.55\textwidth}
         \centering
         \includegraphics[width=\textwidth]{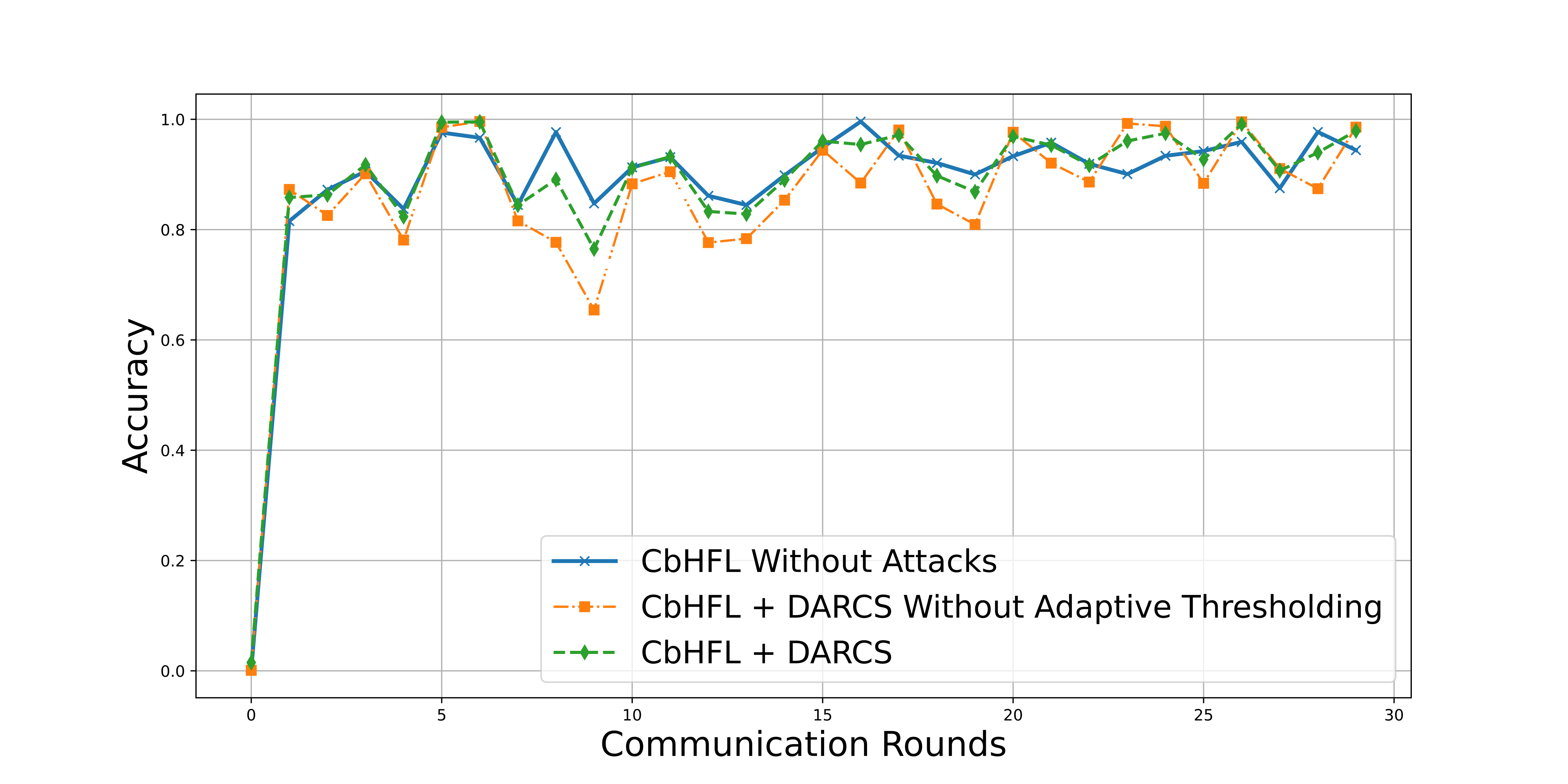}
         \subcaption*{(c)}
         \label{fig: Cont-Gradient and Noise}
     \end{subfigure}
        \caption{Comparison of accuracy at the EPC with and without adaptive thresholding. a) Noise attack with Mean= 2, Variance= 0.3, b) Gradient ascent attack, c) Noise and gradient ascent attack.}
        \label{fig: EPC Accuracy for Count Attack}
\end{figure}
%%==========================FIG 2=============================%%
\begin{figure}[!t]
     \centering
     \begin{subfigure}{0.55\textwidth}
         \centering
         \includegraphics[width=\textwidth]{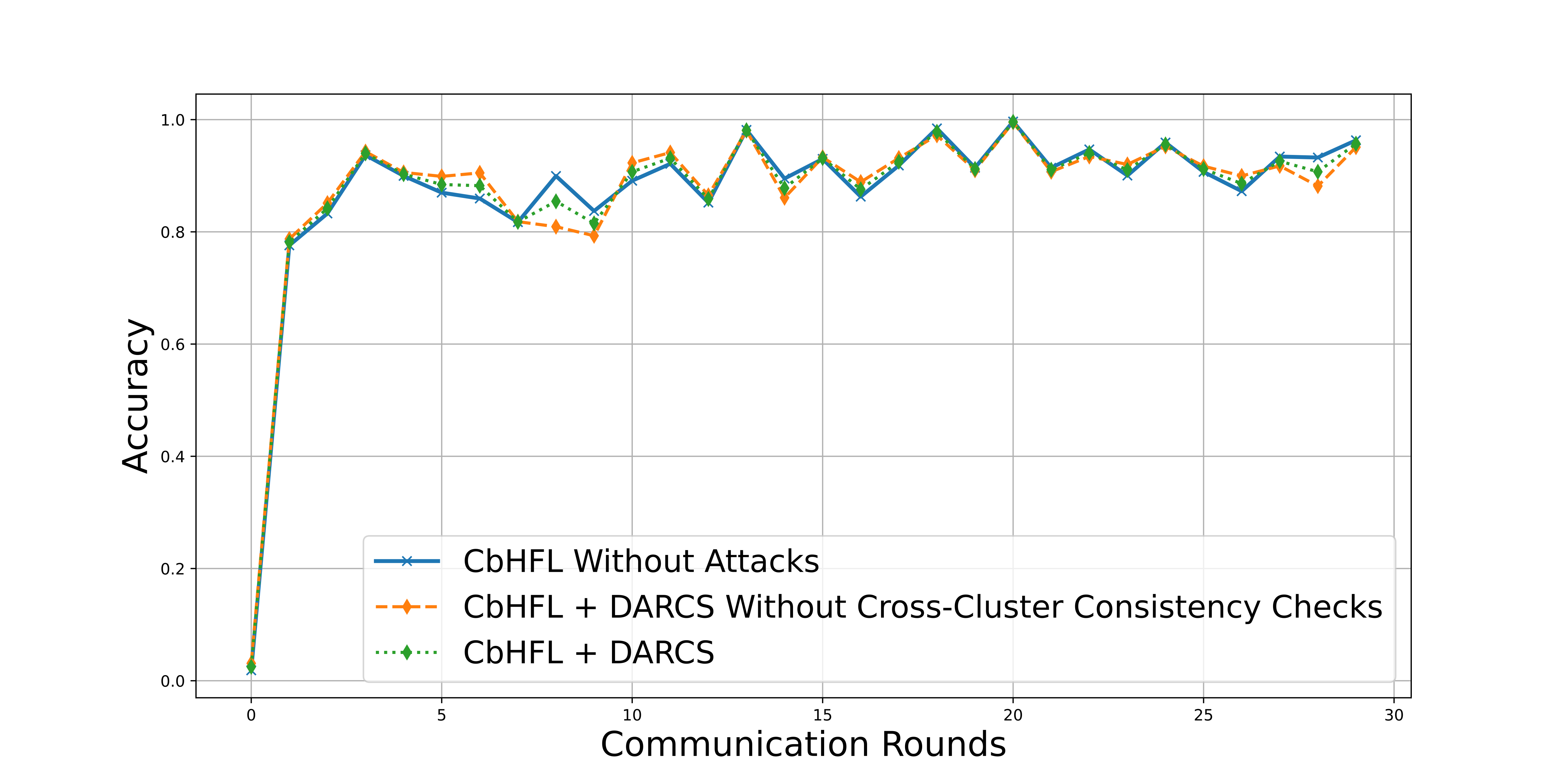}
         \subcaption*{(a)}
         \label{fig: Cont-0-0.3}
     \end{subfigure}
     \vfill
     \begin{subfigure}{0.55\textwidth}
         \centering
         \includegraphics[width=\textwidth]{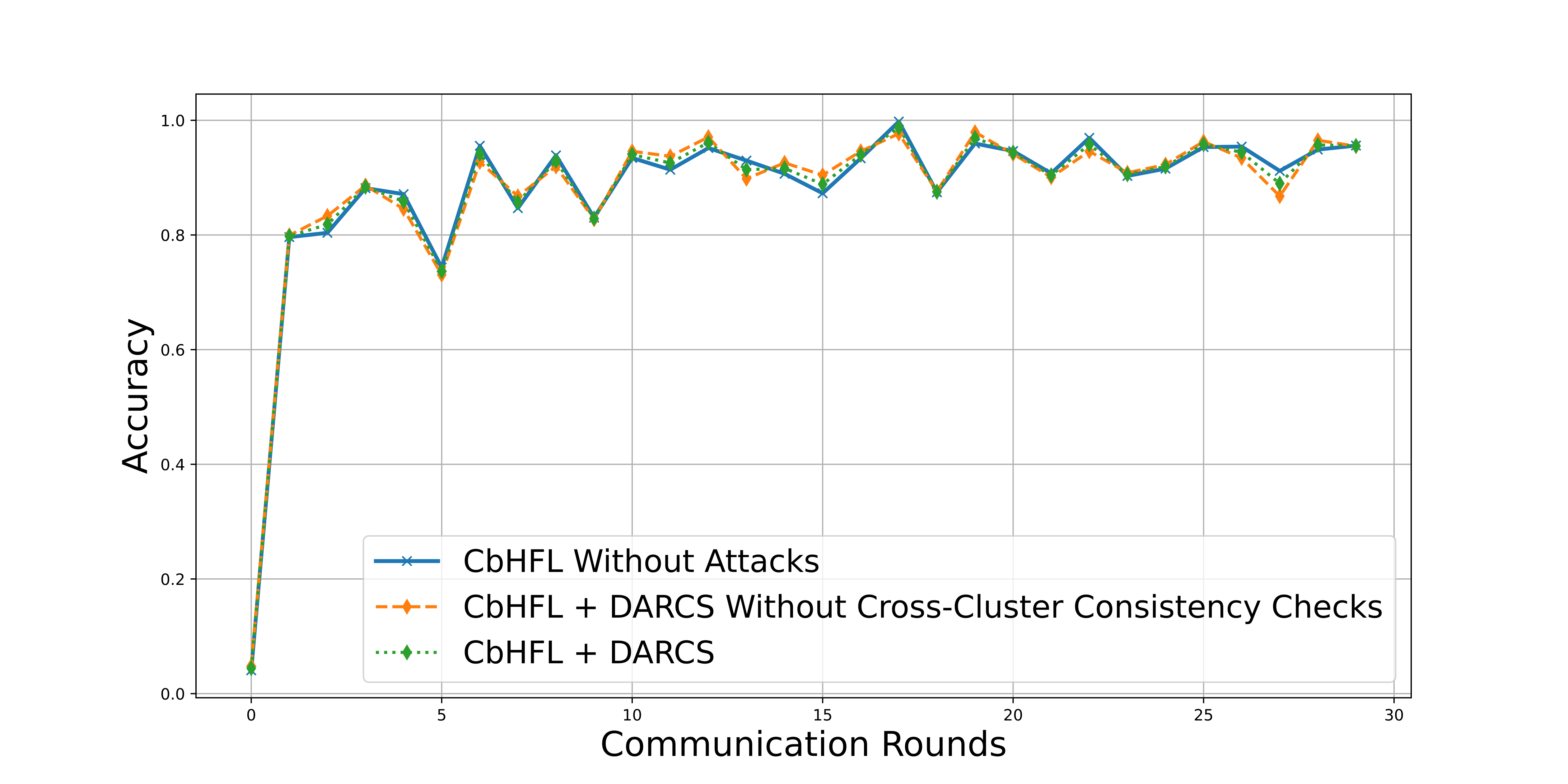}
         \subcaption*{(b)}
         \label{fig: Cont-Gradient}
     \end{subfigure}
     \vfill
     \begin{subfigure}{0.55\textwidth}
         \centering
         \includegraphics[width=\textwidth]{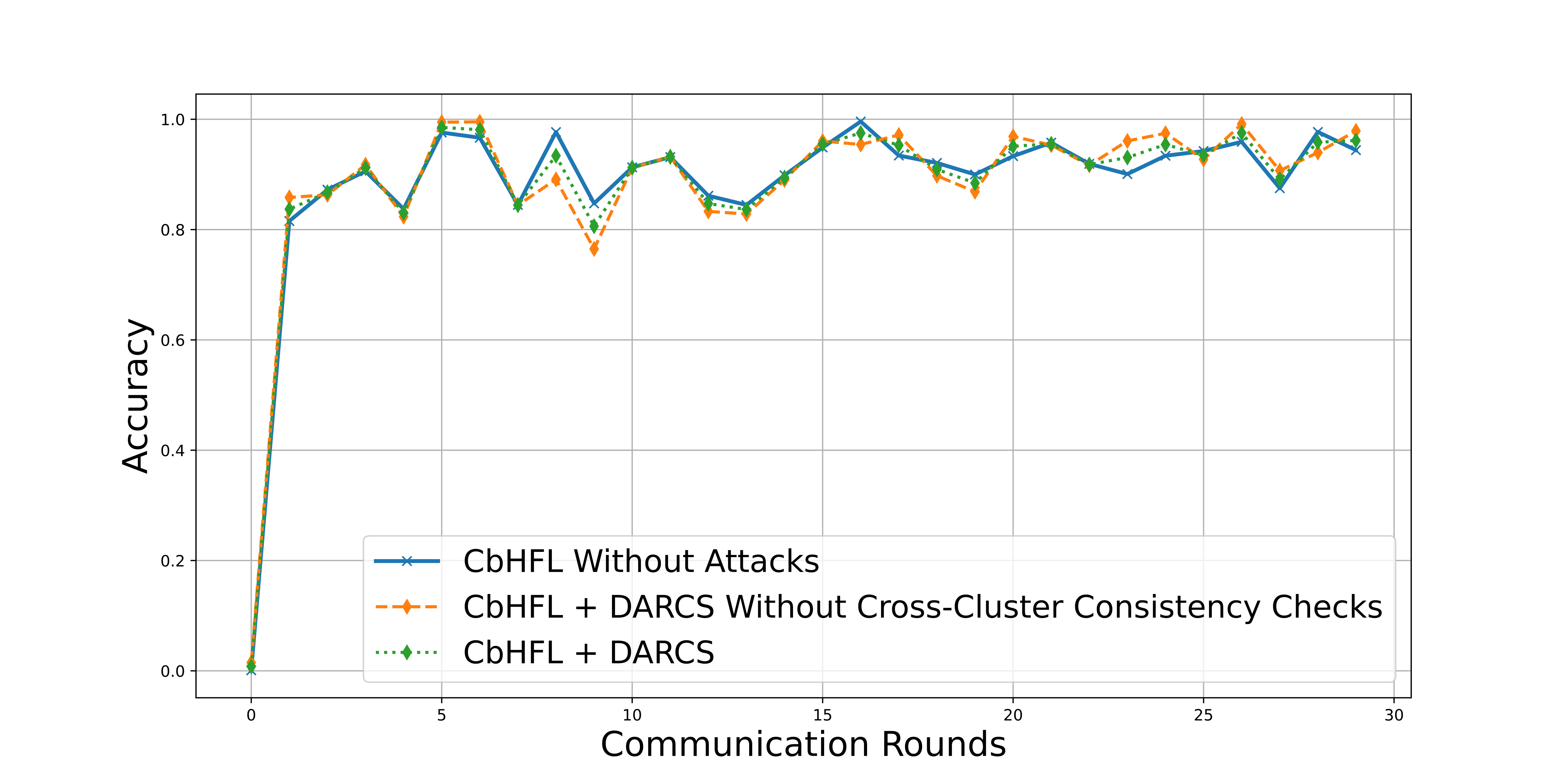}
         \subcaption*{(c)}
         \label{fig: Cont-Gradient and Noise}
     \end{subfigure}
        \caption{Comparison of accuracy at the EPC with and without cross-cluster consistency check. a) Noise attack with Mean= 2, Variance= 0.3, b) Gradient ascent attack, c) Noise and gradient ascent attack.}
        \label{fig: EPC Accuracy for Count Attack}
\end{figure}

In this section, we evaluate the impact of incorporating adaptive thresholding and cross-cluster consistency checks into the proposed algorithm. The goal is to demonstrate their effectiveness and highlight the necessity of integrating both components. The analysis focuses on a scenario with 25 vehicles operating at a 100-meter transmission range in a single-hop configuration. By comparing the performance of the algorithm with and without these features, we provide a detailed assessment of their contributions to the overall robustness and security of the framework.

%%========== Performance Evaluation of adaptive thresholding ==========%%
Figure 1 illustrates the impact of adaptive thresholding on the performance of the proposed algorithm for various types of attacks: high noise, gradient ascent, and a combination of both. Under Gaussian noise attack, static thresholding exhibits an accuracy drop of roughly 5\% from the 95\% baseline, while adaptive thresholding shows a less significant decrease of around 1\%. The sharp dip under static thresholding, reaching as low as 75\%, is completely mitigated by adaptive thresholding. Under gradient ascent attacks, static thresholding results in a loss of approximately 6\%, whereas adaptive thresholding restricts the decline to nearly 2\%. Even in the most severe combined noise and gradient ascent scenario, static thresholding causes a drop of almost 7\%, while adaptive thresholding confines the loss to approximately 3\%. These results clearly demonstrate that static thresholds allow performance degradation of up to 7\%, whereas adaptive thresholding, by leveraging each vehicle’s historical reliability, limits accuracy loss to within 1–3\% of the baseline, thereby preserving overall system stability under all attack types. 
%In summary, adaptive thresholding significantly improves resilience, particularly under harsh conditions involving noise or combined attacks. In contrast, static thresholding exhibits marked fluctuations after the onset of attacks, due to its inability to adjust to changing conditions. By incorporating historical performance information, adaptive thresholding effectively detects and mitigates vehicles transitioning from benign to malicious behavior, thereby preserving overall system stability.
%When noise is combined with gradient ascent attacks, the critical importance of adaptive thresholding becomes evident in dynamically responding to varying attack intensities.

%%========== Performance Evaluation of cross-cluster consistency check ==========%%
Figure 2 demonstrates the impact of the cross-cluster consistency check on the performance of the proposed algorithm, evaluated under the same scenarios and attack conditions. Incorporating the cross-cluster consistency check enhances system resilience by identifying deviations from the mean average cosine similarity across clusters. If a CH or a cluster's weighted average exhibits malicious behavior significantly deviating from the global model, this anomaly is immediately detected and mitigated by excluding the malicious entity. As illustrated in the figure, under Gaussian noise attack, omitting the cross‑cluster consistency check results in an accuracy drop of roughly 6\% from the 95\% baseline, while incorporating the consistency check limits the decrease to about 3\%. Without the check, the deepest accuracy dip, reaching 80\%, is entirely smoothed out by the consistency check. Under gradient ascent attacks, performance without the consistency check declines by around 5\%, whereas including the check reduces the drop to about 2\%. In the combined noise and gradient ascent scenario, the no-check configuration leads to a degradation of nearly 7\%, while the consistency check confines the loss to approximately 3\%. These results demonstrate that, without cross‑cluster consistency checks, the system is unable to detect coordinated malicious updates, leading up to 7\% accuracy loss. In contrast, verifying inter-cluster alignment through the consistency check limits the degradation to within 2–3\% of the baseline, preserving system stability across all attack types. 
%In conclusion, an algorithm without a cross-cluster consistency check struggles under severe attack conditions, particularly when subjected to a combination of high Gaussian noise and gradient ascent manipulation. In contrast, integrating the cross-cluster consistency check results in performance comparable to a scenario without any attacks, demonstrating its effectiveness in stabilizing the system. 

%These findings demonstrate that adaptive thresholding dynamically adjusts to changing conditions, thereby enhancing algorithmic robustness, and cross-cluster consistency checks reinforce reliability by validating the consistency of updates across different clusters.

%These findings emphasize the critical role of cross-cluster consistency checks in enhancing the system's defense mechanisms, ensuring stability, and aligning its performance with that of the CbHFL framework under attack-free conditions. Furthermore, they highlight the necessity of combining adaptive thresholding, cross-cluster consistency checks, cosine similarity, and Z-Score detection within the system. Together, these components, unified through the reliability metric, form a robust and comprehensive algorithm capable of detecting and mitigating a wide range of attacks, ensuring strong protection from the CH to the EPC.
%%===== Performance Comparison of Proposed Algorithm to Benchmark Algorithms =====%%
\subsection{Performance Comparison of Proposed Algorithm with Benchmark Algorithms}

\begin{figure}[!t]
     \centering
     \begin{subfigure}{0.55\textwidth}
         \centering
         \includegraphics[width=\textwidth]{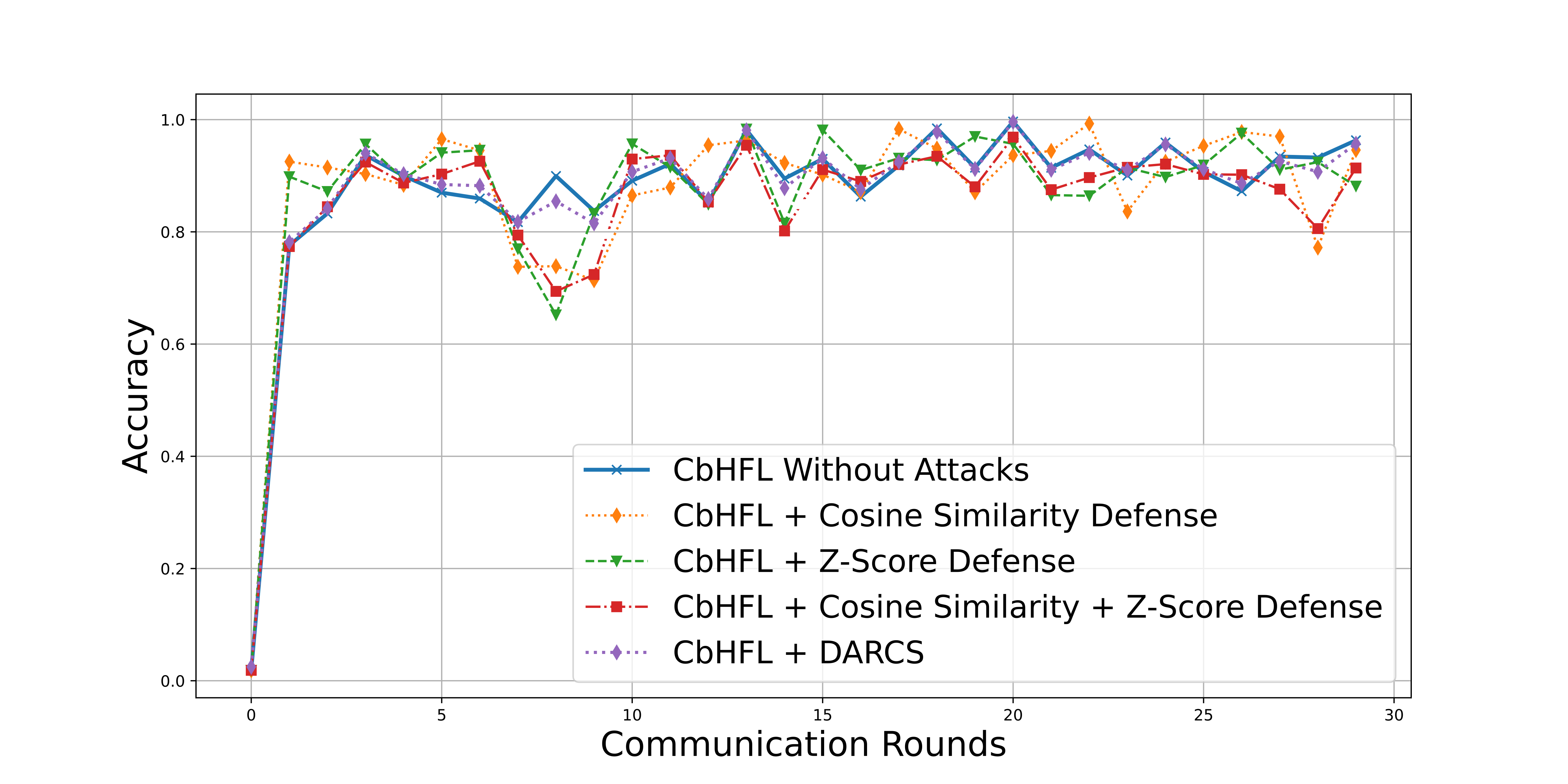}
         \subcaption*{(a)}
         \label{fig: 25-100-1-Noise}
     \end{subfigure}
     \vfill
     \begin{subfigure}{0.55\textwidth}
         \centering
         \includegraphics[width=\textwidth]{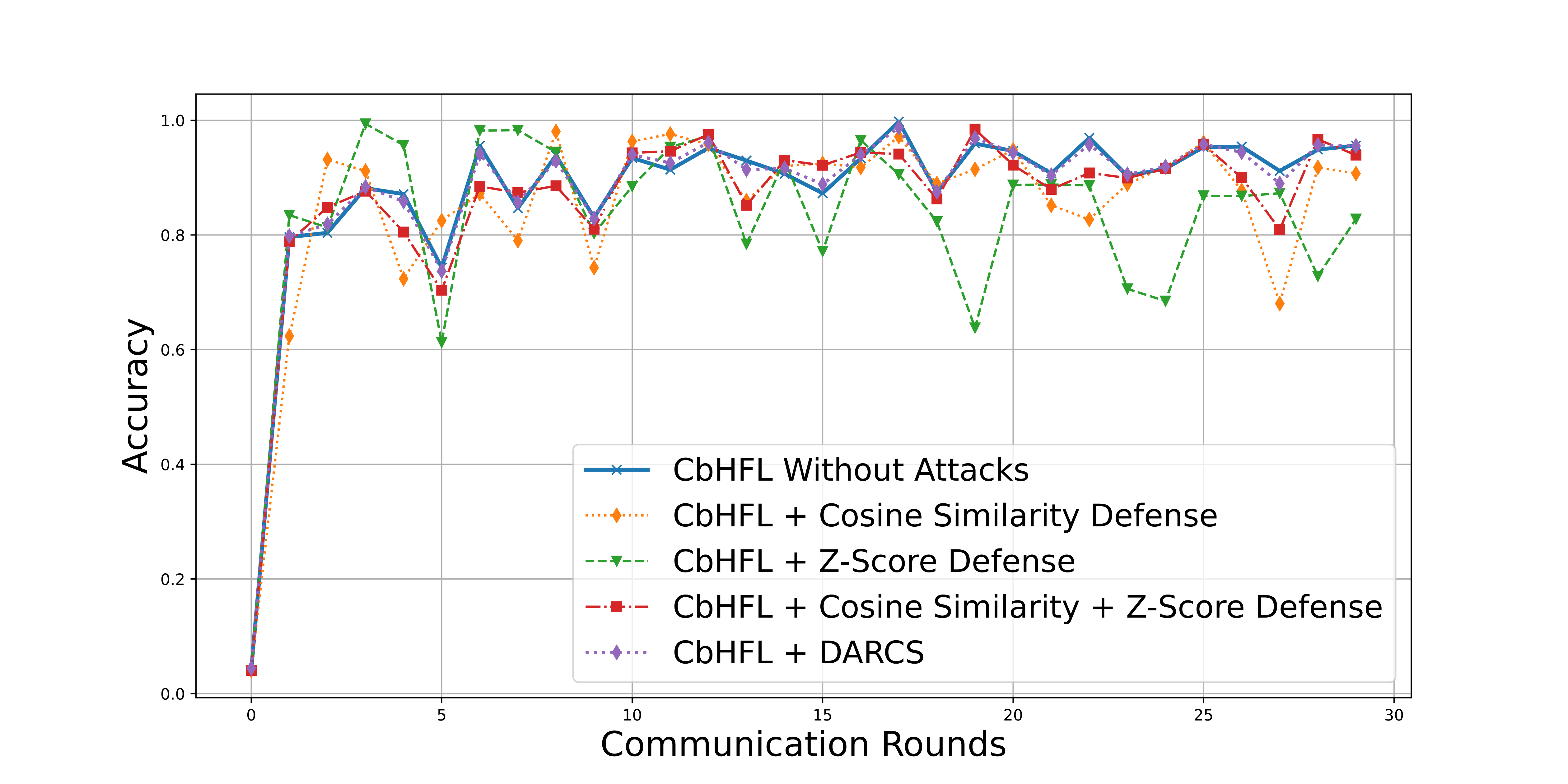}
         \subcaption*{(b)}
         \label{fig: 25-100-1-Gradient}
     \end{subfigure}
     \vfill
     \begin{subfigure}{0.55\textwidth}
         \centering
         \includegraphics[width=\textwidth]{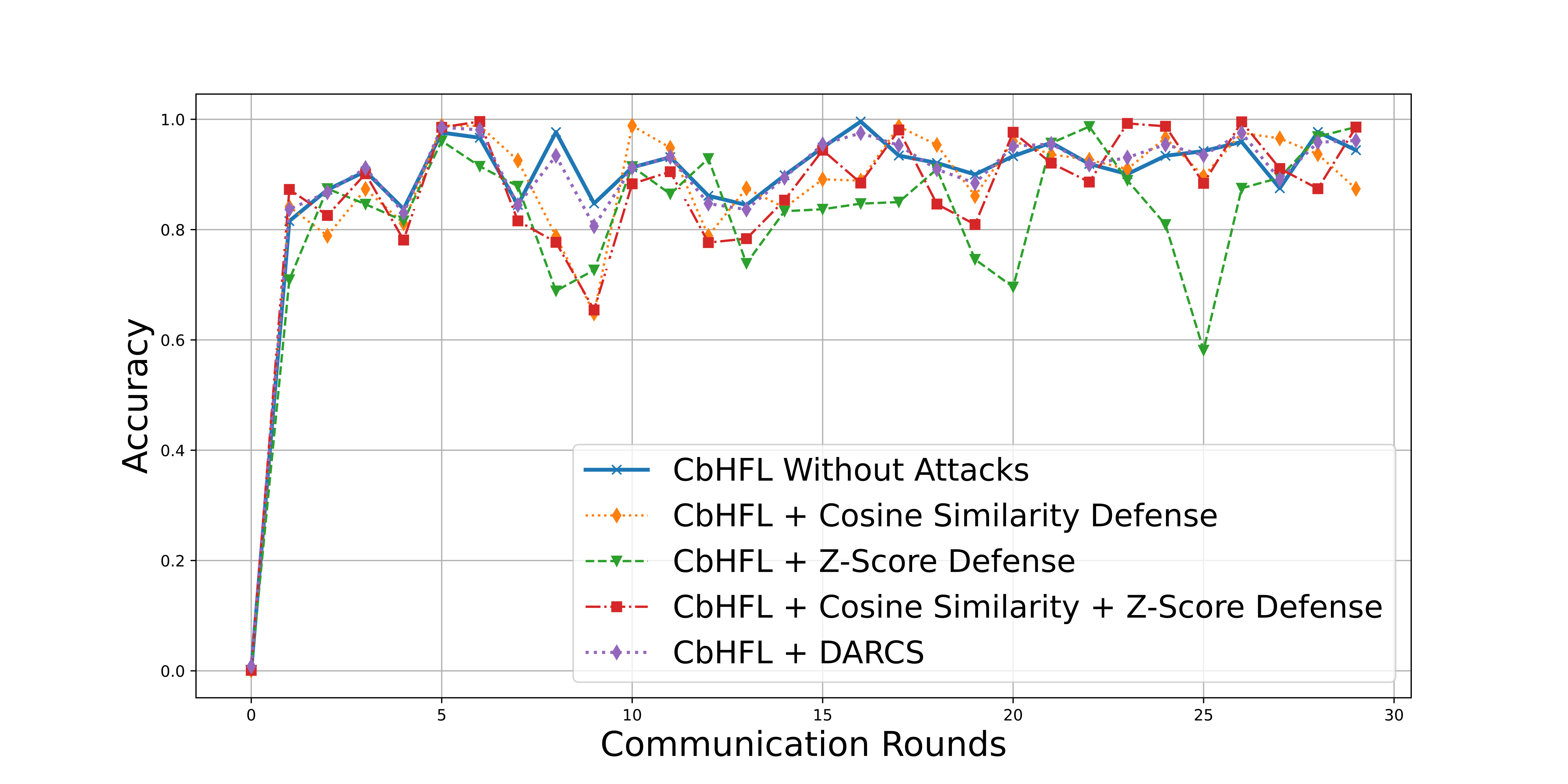}
         \subcaption*{(c)}
         \label{fig: 25-100-1-Gradient and Noise}
     \end{subfigure}
        \caption{Comparison of accuracy at the EPC with benchmark algorithms for 25 vehicles, 100-meter transmission range, and 1-hop communication: a) Noise attack with Mean= 2, Variance= 0.3, b) Gradient ascent attack, c) Noise and gradient ascent attack.}
        \label{fig: EPC Accuracy for 10th Attack}
\end{figure}

Figure 3 shows the accuracy at the EPC under various attack scenarios in a 1‑hop communication setup with 25 vehicles and a 100‑meter transmission range. The baseline performance of CbHFL without any attacks achieves approximately 95\% accuracy, representing the ideal state. Under severe adversarial conditions—particularly when high Gaussian noise is combined with gradient ascent attacks—benchmark methods experience substantial performance degradation, resulting in significant accuracy fluctuations and system instability. Under pure Gaussian noise, the Z score only approach exhibits a drop in accuracy of approximately 12\% from the baseline, while the cosine similarity only method shows a slight decrease of around 10\%. Even when these defenses are combined, the aggregated accuracy remains about 8\% lower than the ideal. Under gradient ascent attacks, the cosine similarity method outperforms the Z‑score approach, with accuracy reductions of approximately 10\% versus 15\%, respectively. When both attack types are compounded, benchmark methods incur losses ranging from 12\% (cosine similarity) to as much as 17\% (Z‑score). In contrast, our proposed algorithm (CbHFL+DARCS) consistently maintains performance very close to the baseline, with only a minimal drop of approximately 2–3\% even under the most challenging conditions. Across all attack types, DARCS delivers the highest resilience, followed by the combined Z‑score and cosine similarity defense, followed by the cosine similarity‑only method, with Z‑score‑only performing worst. Under pure Gaussian noise, our proposed algorithm best mitigates random perturbations, while Z‑score‑only suffers the greatest degradation. In the gradient ascent scenario, DARCS again leads, the combined defense comes second, and Z‑score‑only remains the weakest. When attacks are compounded, DARCS still holds within 3\% of the baseline, the combined defense offers partial mitigation, and Z‑score‑only incurs the most severe drop. Moreover, similar trends are observed in a 3‑hop environment, though performance degradation in benchmark techniques is even more pronounced due to reduced inter‑cluster diversity in larger clusters. Overall, these results clearly demonstrate that while conventional benchmark techniques provide only partial protection, our DARCS algorithm delivers superior robustness and stability, achieving performance within just 2–3\% of the baseline, thus offering a highly reliable solution for dynamic vehicular HFL environments.
\begin{table*}[ht]
\centering
\caption{Convergence time under varying attack types, number of vehicles, transmission ranges, and convergence threshold $\epsilon$ for 1-hop communication.}
\begin{tabular}{|ccccccc|}
\hline
\multicolumn{7}{|c|}{25 Vehicles, Tx Range = 100 Meters}                                                                                                                                                                                                                                                                                                                                                                                                                                                                                                                                                                          \\ \hline
\multicolumn{2}{|c|}{}                                                                                                                                            & \multicolumn{1}{c|}{\begin{tabular}[c]{@{}c@{}}CbHFL \\ Without Attacks\end{tabular}} & \multicolumn{1}{c|}{\begin{tabular}[c]{@{}c@{}}CbHFL +\\  Cosine Similarity Defense\end{tabular}} & \multicolumn{1}{c|}{\begin{tabular}[c]{@{}c@{}}CbHFL + \\ Z-Score Defense\end{tabular}} & \multicolumn{1}{c|}{\begin{tabular}[c]{@{}c@{}}CbHFL + \\ Cosine Similarity + \\ Z-Score Defense\end{tabular}} & \begin{tabular}[c]{@{}c@{}}CbHFL + \\ DARCS\end{tabular} \\ \hline
\multicolumn{1}{|c|}{\multirow{3}{*}{\begin{tabular}[c]{@{}c@{}}Noise Attack with\\ Mean = 2,\\ Var = 0.3\end{tabular}}} & \multicolumn{1}{c|}{$\epsilon= 0.01$}  & \multicolumn{1}{c|}{61}                                                               & \multicolumn{1}{c|}{72}                                                                           & \multicolumn{1}{c|}{76}                                                                 & \multicolumn{1}{c|}{69}                                                                                        & 63                                                     \\ \cline{2-7} 
\multicolumn{1}{|c|}{}                                                                                                   & \multicolumn{1}{c|}{$\epsilon= 0.005$} & \multicolumn{1}{c|}{65}                                                               & \multicolumn{1}{c|}{75}                                                                           & \multicolumn{1}{c|}{$\infty$}                                                           & \multicolumn{1}{c|}{74}                                                                                        & 68                                                     \\ \cline{2-7} 
\multicolumn{1}{|c|}{}                                                                                                   & \multicolumn{1}{c|}{$\epsilon= 0.001$} & \multicolumn{1}{c|}{73}                                                               & \multicolumn{1}{c|}{$\infty$}                                                                     & \multicolumn{1}{c|}{$\infty$}                                                           & \multicolumn{1}{c|}{$\infty$}                                                                                  & 74                                                     \\ \hline
\multicolumn{1}{|c|}{\multirow{3}{*}{Gradient Ascent Attack}}                                                            & \multicolumn{1}{c|}{$\epsilon= 0.01$}  & \multicolumn{1}{c|}{53}                                                               & \multicolumn{1}{c|}{68}                                                                           & \multicolumn{1}{c|}{74}                                                                 & \multicolumn{1}{c|}{57}                                                                                        & 55                                                     \\ \cline{2-7} 
\multicolumn{1}{|c|}{}                                                                                                   & \multicolumn{1}{c|}{$\epsilon= 0.005$} & \multicolumn{1}{c|}{58}                                                               & \multicolumn{1}{c|}{73}                                                                           & \multicolumn{1}{c|}{$\infty$}                                                           & \multicolumn{1}{c|}{68}                                                                                        & 61                                                     \\ \cline{2-7} 
\multicolumn{1}{|c|}{}                                                                                                   & \multicolumn{1}{c|}{$\epsilon= 0.001$} & \multicolumn{1}{c|}{69}                                                               & \multicolumn{1}{c|}{74}                                                                           & \multicolumn{1}{c|}{$\infty$}                                                           & \multicolumn{1}{c|}{74}                                                                                        & 72                                                     \\ \hline
\multicolumn{1}{|c|}{\multirow{3}{*}{Noise + Gradient Ascent Attack}}                                                    & \multicolumn{1}{c|}{$\epsilon= 0.01$}  & \multicolumn{1}{c|}{62}                                                               & \multicolumn{1}{c|}{74}                                                                           & \multicolumn{1}{c|}{$\infty$}                                                           & \multicolumn{1}{c|}{70}                                                                                        & 65                                                     \\ \cline{2-7} 
\multicolumn{1}{|c|}{}                                                                                                   & \multicolumn{1}{c|}{$\epsilon= 0.005$} & \multicolumn{1}{c|}{70}                                                               & \multicolumn{1}{c|}{$\infty$}                                                                     & \multicolumn{1}{c|}{$\infty$}                                                           & \multicolumn{1}{c|}{74}                                                                                        & 73                                                     \\ \cline{2-7} 
\multicolumn{1}{|c|}{}                                                                                                   & \multicolumn{1}{c|}{$\epsilon= 0.001$} & \multicolumn{1}{c|}{74}                                                               & \multicolumn{1}{c|}{$\infty$}                                                                     & \multicolumn{1}{c|}{$\infty$}                                                           & \multicolumn{1}{c|}{$\infty$}                                                                                  & 74                                                     \\ \hline
\multicolumn{7}{|c|}{25 Vehicles, Tx Range = 500 Meters}                                                                                                                                                                                                                                                                                                                                                                                                                                                                                                                                                                          \\ \hline
\multicolumn{1}{|c|}{\multirow{3}{*}{\begin{tabular}[c]{@{}c@{}}Noise Attack with\\ Mean = 2,\\ Var = 0.3\end{tabular}}} & \multicolumn{1}{c|}{$\epsilon= 0.01$}  & \multicolumn{1}{c|}{59}                                                               & \multicolumn{1}{c|}{66}                                                                           & \multicolumn{1}{c|}{71}                                                                 & \multicolumn{1}{c|}{67}                                                                                        & 59                                                     \\ \cline{2-7} 
\multicolumn{1}{|c|}{}                                                                                                   & \multicolumn{1}{c|}{$\epsilon= 0.005$} & \multicolumn{1}{c|}{58}                                                               & \multicolumn{1}{c|}{70}                                                                           & \multicolumn{1}{c|}{$\infty$}                                                           & \multicolumn{1}{c|}{68}                                                                                        & 63                                                     \\ \cline{2-7} 
\multicolumn{1}{|c|}{}                                                                                                   & \multicolumn{1}{c|}{$\epsilon= 0.001$} & \multicolumn{1}{c|}{65}                                                               & \multicolumn{1}{c|}{$\infty$}                                                                     & \multicolumn{1}{c|}{$\infty$}                                                           & \multicolumn{1}{c|}{$\infty$}                                                                                  & 68                                                     \\ \hline
\multicolumn{1}{|c|}{\multirow{3}{*}{Gradient Ascent Attack}}                                                            & \multicolumn{1}{c|}{$\epsilon= 0.01$}  & \multicolumn{1}{c|}{47}                                                               & \multicolumn{1}{c|}{62}                                                                           & \multicolumn{1}{c|}{68}                                                                 & \multicolumn{1}{c|}{52}                                                                                        & 49                                                     \\ \cline{2-7} 
\multicolumn{1}{|c|}{}                                                                                                   & \multicolumn{1}{c|}{$\epsilon= 0.005$} & \multicolumn{1}{c|}{55}                                                               & \multicolumn{1}{c|}{67}                                                                           & \multicolumn{1}{c|}{75}                                                                 & \multicolumn{1}{c|}{62}                                                                                        & 58                                                     \\ \cline{2-7} 
\multicolumn{1}{|c|}{}                                                                                                   & \multicolumn{1}{c|}{$\epsilon= 0.001$} & \multicolumn{1}{c|}{64}                                                               & \multicolumn{1}{c|}{70}                                                                           & \multicolumn{1}{c|}{$\infty$}                                                           & \multicolumn{1}{c|}{69}                                                                                        & 69                                                     \\ \hline
\multicolumn{1}{|c|}{\multirow{3}{*}{Noise + Gradient Ascent Attack}}                                                    & \multicolumn{1}{c|}{$\epsilon= 0.01$}  & \multicolumn{1}{c|}{58}                                                               & \multicolumn{1}{c|}{68}                                                                           & \multicolumn{1}{c|}{70}                                                                 & \multicolumn{1}{c|}{65}                                                                                        & 61                                                     \\ \cline{2-7} 
\multicolumn{1}{|c|}{}                                                                                                   & \multicolumn{1}{c|}{$\epsilon= 0.005$} & \multicolumn{1}{c|}{66}                                                               & \multicolumn{1}{c|}{73}                                                                           & \multicolumn{1}{c|}{$\infty$}                                                           & \multicolumn{1}{c|}{71}                                                                                        & 66                                                     \\ \cline{2-7} 
\multicolumn{1}{|c|}{}                                                                                                   & \multicolumn{1}{c|}{$\epsilon= 0.001$} & \multicolumn{1}{c|}{69}                                                               & \multicolumn{1}{c|}{$\infty$}                                                                     & \multicolumn{1}{c|}{$\infty$}                                                           & \multicolumn{1}{c|}{74}                                                                                        & 69                                                     \\ \hline
\multicolumn{7}{|c|}{50 Vehicles, Tx Range = 100 Meters}                                                                                                                                                                                                                                                                                                                                                                                                                                                                                                                                                                          \\ \hline
\multicolumn{1}{|c|}{\multirow{3}{*}{\begin{tabular}[c]{@{}c@{}}Noise Attack with\\ Mean = 2,\\ Var = 0.3\end{tabular}}} & \multicolumn{1}{c|}{$\epsilon= 0.01$}  & \multicolumn{1}{c|}{68}                                                               & \multicolumn{1}{c|}{80}                                                                           & \multicolumn{1}{c|}{81}                                                                 & \multicolumn{1}{c|}{75}                                                                                        & 71                                                     \\ \cline{2-7} 
\multicolumn{1}{|c|}{}                                                                                                   & \multicolumn{1}{c|}{$\epsilon= 0.005$} & \multicolumn{1}{c|}{73}                                                               & \multicolumn{1}{c|}{81}                                                                           & \multicolumn{1}{c|}{$\infty$}                                                           & \multicolumn{1}{c|}{82}                                                                                        & 75                                                     \\ \cline{2-7} 
\multicolumn{1}{|c|}{}                                                                                                   & \multicolumn{1}{c|}{$\epsilon= 0.001$} & \multicolumn{1}{c|}{77}                                                               & \multicolumn{1}{c|}{$\infty$}                                                                     & \multicolumn{1}{c|}{$\infty$}                                                           & \multicolumn{1}{c|}{$\infty$}                                                                                  & 80                                                     \\ \hline
\multicolumn{1}{|c|}{\multirow{3}{*}{Gradient Ascent Attack}}                                                            & \multicolumn{1}{c|}{$\epsilon= 0.01$}  & \multicolumn{1}{c|}{59}                                                               & \multicolumn{1}{c|}{75}                                                                           & \multicolumn{1}{c|}{80}                                                                 & \multicolumn{1}{c|}{63}                                                                                        & 59                                                     \\ \cline{2-7} 
\multicolumn{1}{|c|}{}                                                                                                   & \multicolumn{1}{c|}{$\epsilon= 0.005$} & \multicolumn{1}{c|}{62}                                                               & \multicolumn{1}{c|}{75}                                                                           & \multicolumn{1}{c|}{$\infty$}                                                           & \multicolumn{1}{c|}{74}                                                                                        & 66                                                     \\ \cline{2-7} 
\multicolumn{1}{|c|}{}                                                                                                   & \multicolumn{1}{c|}{$\epsilon= 0.001$} & \multicolumn{1}{c|}{72}                                                               & \multicolumn{1}{c|}{81}                                                                           & \multicolumn{1}{c|}{$\infty$}                                                           & \multicolumn{1}{c|}{78}                                                                                        & 77                                                     \\ \hline
\multicolumn{1}{|c|}{\multirow{3}{*}{Noise + Gradient Ascent Attack}}                                                    & \multicolumn{1}{c|}{$\epsilon= 0.01$}  & \multicolumn{1}{c|}{68}                                                               & \multicolumn{1}{c|}{77}                                                                           & \multicolumn{1}{c|}{$\infty$}                                                           & \multicolumn{1}{c|}{76}                                                                                        & 71                                                     \\ \cline{2-7} 
\multicolumn{1}{|c|}{}                                                                                                   & \multicolumn{1}{c|}{$\epsilon= 0.005$} & \multicolumn{1}{c|}{73}                                                               & \multicolumn{1}{c|}{$\infty$}                                                                     & \multicolumn{1}{c|}{$\infty$}                                                           & \multicolumn{1}{c|}{81}                                                                                        & 77                                                     \\ \cline{2-7} 
\multicolumn{1}{|c|}{}                                                                                                   & \multicolumn{1}{c|}{$\epsilon= 0.001$} & \multicolumn{1}{c|}{77}                                                               & \multicolumn{1}{c|}{$\infty$}                                                                     & \multicolumn{1}{c|}{$\infty$}                                                           & \multicolumn{1}{c|}{$\infty$}                                                                                  & 78                                                     \\ \hline
\end{tabular}
\end{table*}
%%==========================TABLE II=============================%%
% \usepackage{multirow}
\begin{table*}[ht]
\centering
\caption{Convergence time under varying attack types, number of vehicles, transmission ranges, and convergence threshold $\epsilon$ for 3-hop communication.}
\begin{tabular}{|ccccccc|}
\hline
\multicolumn{7}{|c|}{25 Vehicles, Tx Range = 100 Meters}                                                                                                                                                                                                                                                                                                                                                                                                                                                                                                                                                                            \\ \hline
\multicolumn{2}{|c|}{}                                                                                                                                            & \multicolumn{1}{c|}{\begin{tabular}[c]{@{}c@{}}CbHFL \\ Without Attacks\end{tabular}} & \multicolumn{1}{c|}{\begin{tabular}[c]{@{}c@{}}CbHFL +\\  Cosine Similarity Defense\end{tabular}} & \multicolumn{1}{c|}{\begin{tabular}[c]{@{}c@{}}CbHFL + \\ Z-Score Defense\end{tabular}} & \multicolumn{1}{c|}{\begin{tabular}[c]{@{}c@{}}CbHFL + \\ Cosine Similarity + \\ Z-Score Defense\end{tabular}} & \begin{tabular}[c]{@{}c@{}}CbHFL + \\ DARCS\end{tabular} \\ \hline
\multicolumn{1}{|c|}{\multirow{3}{*}{\begin{tabular}[c]{@{}c@{}}Noise Attack with\\ Mean = 2,\\ Var = 0.3\end{tabular}}} & \multicolumn{1}{c|}{$\epsilon= 0.01$}  & \multicolumn{1}{c|}{60}                                                               & \multicolumn{1}{c|}{70}                                                                           & \multicolumn{1}{c|}{75}                                                                 & \multicolumn{1}{c|}{68}                                                                                        & 60                                                       \\ \cline{2-7} 
\multicolumn{1}{|c|}{}                                                                                                   & \multicolumn{1}{c|}{$\epsilon= 0.005$} & \multicolumn{1}{c|}{62}                                                               & \multicolumn{1}{c|}{73}                                                                           & \multicolumn{1}{c|}{$\infty$}                                                           & \multicolumn{1}{c|}{70}                                                                                        & 66                                                       \\ \cline{2-7} 
\multicolumn{1}{|c|}{}                                                                                                   & \multicolumn{1}{c|}{$\epsilon= 0.001$} & \multicolumn{1}{c|}{69}                                                               & \multicolumn{1}{c|}{$\infty$}                                                                     & \multicolumn{1}{c|}{$\infty$}                                                           & \multicolumn{1}{c|}{$\infty$}                                                                                  & 72                                                       \\ \hline
\multicolumn{1}{|c|}{\multirow{3}{*}{Gradient Ascent Attack}}                                                            & \multicolumn{1}{c|}{$\epsilon= 0.01$}  & \multicolumn{1}{c|}{51}                                                               & \multicolumn{1}{c|}{66}                                                                           & \multicolumn{1}{c|}{71}                                                                 & \multicolumn{1}{c|}{54}                                                                                        & 51                                                       \\ \cline{2-7} 
\multicolumn{1}{|c|}{}                                                                                                   & \multicolumn{1}{c|}{$\epsilon= 0.005$} & \multicolumn{1}{c|}{56}                                                               & \multicolumn{1}{c|}{69}                                                                           & \multicolumn{1}{c|}{$\infty$}                                                           & \multicolumn{1}{c|}{66}                                                                                        & 60                                                       \\ \cline{2-7} 
\multicolumn{1}{|c|}{}                                                                                                   & \multicolumn{1}{c|}{$\epsilon= 0.001$} & \multicolumn{1}{c|}{66}                                                               & \multicolumn{1}{c|}{71}                                                                           & \multicolumn{1}{c|}{$\infty$}                                                           & \multicolumn{1}{c|}{74}                                                                                        & 71                                                       \\ \hline
\multicolumn{1}{|c|}{\multirow{3}{*}{Noise + Gradient Ascent Attack}}                                                    & \multicolumn{1}{c|}{$\epsilon= 0.01$}  & \multicolumn{1}{c|}{59}                                                               & \multicolumn{1}{c|}{70}                                                                           & \multicolumn{1}{c|}{$\infty$}                                                           & \multicolumn{1}{c|}{67}                                                                                        & 63                                                       \\ \cline{2-7} 
\multicolumn{1}{|c|}{}                                                                                                   & \multicolumn{1}{c|}{$\epsilon= 0.005$} & \multicolumn{1}{c|}{69}                                                               & \multicolumn{1}{c|}{$\infty$}                                                                     & \multicolumn{1}{c|}{$\infty$}                                                           & \multicolumn{1}{c|}{73}                                                                                        & 69                                                       \\ \cline{2-7} 
\multicolumn{1}{|c|}{}                                                                                                   & \multicolumn{1}{c|}{$\epsilon= 0.001$} & \multicolumn{1}{c|}{71}                                                               & \multicolumn{1}{c|}{$\infty$}                                                                     & \multicolumn{1}{c|}{$\infty$}                                                           & \multicolumn{1}{c|}{$\infty$}                                                                                  & 72                                                       \\ \hline
\multicolumn{7}{|c|}{25 Vehicles, Tx Range = 500 Meters}                                                                                                                                                                                                                                                                                                                                                                                                                                                                                                                                                                            \\ \hline
\multicolumn{1}{|c|}{\multirow{3}{*}{\begin{tabular}[c]{@{}c@{}}Noise Attack with\\ Mean = 2,\\ Var = 0.3\end{tabular}}} & \multicolumn{1}{c|}{$\epsilon= 0.01$}  & \multicolumn{1}{c|}{58}                                                               & \multicolumn{1}{c|}{62}                                                                           & \multicolumn{1}{c|}{71}                                                                 & \multicolumn{1}{c|}{62}                                                                                        & 59                                                       \\ \cline{2-7} 
\multicolumn{1}{|c|}{}                                                                                                   & \multicolumn{1}{c|}{$\epsilon= 0.005$} & \multicolumn{1}{c|}{56}                                                               & \multicolumn{1}{c|}{68}                                                                           & \multicolumn{1}{c|}{$\infty$}                                                           & \multicolumn{1}{c|}{66}                                                                                        & 59                                                       \\ \cline{2-7} 
\multicolumn{1}{|c|}{}                                                                                                   & \multicolumn{1}{c|}{$\epsilon= 0.001$} & \multicolumn{1}{c|}{67}                                                               & \multicolumn{1}{c|}{71}                                                                           & \multicolumn{1}{c|}{$\infty$}                                                           & \multicolumn{1}{c|}{70}                                                                                        & 67                                                       \\ \hline
\multicolumn{1}{|c|}{\multirow{3}{*}{Gradient Ascent Attack}}                                                            & \multicolumn{1}{c|}{$\epsilon= 0.01$}  & \multicolumn{1}{c|}{46}                                                               & \multicolumn{1}{c|}{60}                                                                           & \multicolumn{1}{c|}{65}                                                                 & \multicolumn{1}{c|}{49}                                                                                        & 48                                                       \\ \cline{2-7} 
\multicolumn{1}{|c|}{}                                                                                                   & \multicolumn{1}{c|}{$\epsilon= 0.005$} & \multicolumn{1}{c|}{48}                                                               & \multicolumn{1}{c|}{65}                                                                           & \multicolumn{1}{c|}{68}                                                                 & \multicolumn{1}{c|}{62}                                                                                        & 54                                                       \\ \cline{2-7} 
\multicolumn{1}{|c|}{}                                                                                                   & \multicolumn{1}{c|}{$\epsilon= 0.001$} & \multicolumn{1}{c|}{62}                                                               & \multicolumn{1}{c|}{66}                                                                           & \multicolumn{1}{c|}{$\infty$}                                                           & \multicolumn{1}{c|}{68}                                                                                        & 64                                                       \\ \hline
\multicolumn{1}{|c|}{\multirow{3}{*}{Noise + Gradient Ascent Attack}}                                                    & \multicolumn{1}{c|}{$\epsilon= 0.01$}  & \multicolumn{1}{c|}{53}                                                               & \multicolumn{1}{c|}{65}                                                                           & \multicolumn{1}{c|}{69}                                                                 & \multicolumn{1}{c|}{61}                                                                                        & 57                                                       \\ \cline{2-7} 
\multicolumn{1}{|c|}{}                                                                                                   & \multicolumn{1}{c|}{$\epsilon= 0.005$} & \multicolumn{1}{c|}{64}                                                               & \multicolumn{1}{c|}{70}                                                                           & \multicolumn{1}{c|}{$\infty$}                                                           & \multicolumn{1}{c|}{66}                                                                                        & 65                                                       \\ \cline{2-7} 
\multicolumn{1}{|c|}{}                                                                                                   & \multicolumn{1}{c|}{$\epsilon= 0.001$} & \multicolumn{1}{c|}{66}                                                               & \multicolumn{1}{c|}{74}                                                                           & \multicolumn{1}{c|}{$\infty$}                                                           & \multicolumn{1}{c|}{72}                                                                                        & 69                                                       \\ \hline
\multicolumn{7}{|c|}{50 Vehicles, Tx Range = 100 Meters}                                                                                                                                                                                                                                                                                                                                                                                                                                                                                                                                                                            \\ \hline
\multicolumn{1}{|c|}{\multirow{3}{*}{\begin{tabular}[c]{@{}c@{}}Noise Attack with\\ Mean = 2,\\ Var = 0.3\end{tabular}}} & \multicolumn{1}{c|}{$\epsilon= 0.01$}  & \multicolumn{1}{c|}{64}                                                               & \multicolumn{1}{c|}{76}                                                                           & \multicolumn{1}{c|}{77}                                                                 & \multicolumn{1}{c|}{70}                                                                                        & 65                                                       \\ \cline{2-7} 
\multicolumn{1}{|c|}{}                                                                                                   & \multicolumn{1}{c|}{$\epsilon= 0.005$} & \multicolumn{1}{c|}{67}                                                               & \multicolumn{1}{c|}{77}                                                                           & \multicolumn{1}{c|}{$\infty$}                                                           & \multicolumn{1}{c|}{75}                                                                                        & 69                                                       \\ \cline{2-7} 
\multicolumn{1}{|c|}{}                                                                                                   & \multicolumn{1}{c|}{$\epsilon= 0.001$} & \multicolumn{1}{c|}{76}                                                               & \multicolumn{1}{c|}{$\infty$}                                                                     & \multicolumn{1}{c|}{$\infty$}                                                           & \multicolumn{1}{c|}{$\infty$}                                                                                  & 76                                                       \\ \hline
\multicolumn{1}{|c|}{\multirow{3}{*}{Gradient Ascent Attack}}                                                            & \multicolumn{1}{c|}{$\epsilon= 0.01$}  & \multicolumn{1}{c|}{57}                                                               & \multicolumn{1}{c|}{71}                                                                           & \multicolumn{1}{c|}{75}                                                                 & \multicolumn{1}{c|}{60}                                                                                        & 59                                                       \\ \cline{2-7} 
\multicolumn{1}{|c|}{}                                                                                                   & \multicolumn{1}{c|}{$\epsilon= 0.005$} & \multicolumn{1}{c|}{61}                                                               & \multicolumn{1}{c|}{75}                                                                           & \multicolumn{1}{c|}{$\infty$}                                                           & \multicolumn{1}{c|}{70}                                                                                        & 62                                                       \\ \cline{2-7} 
\multicolumn{1}{|c|}{}                                                                                                   & \multicolumn{1}{c|}{$\epsilon= 0.001$} & \multicolumn{1}{c|}{71}                                                               & \multicolumn{1}{c|}{77}                                                                           & \multicolumn{1}{c|}{$\infty$}                                                           & \multicolumn{1}{c|}{75}                                                                                        & 73                                                       \\ \hline
\multicolumn{1}{|c|}{\multirow{3}{*}{Noise + Gradient Ascent Attack}}                                                    & \multicolumn{1}{c|}{$\epsilon= 0.01$}  & \multicolumn{1}{c|}{66}                                                               & \multicolumn{1}{c|}{78}                                                                           & \multicolumn{1}{c|}{$\infty$}                                                           & \multicolumn{1}{c|}{72}                                                                                        & 66                                                       \\ \cline{2-7} 
\multicolumn{1}{|c|}{}                                                                                                   & \multicolumn{1}{c|}{$\epsilon= 0.005$} & \multicolumn{1}{c|}{71}                                                               & \multicolumn{1}{c|}{$\infty$}                                                                     & \multicolumn{1}{c|}{$\infty$}                                                           & \multicolumn{1}{c|}{78}                                                                                        & 74                                                       \\ \cline{2-7} 
\multicolumn{1}{|c|}{}                                                                                                   & \multicolumn{1}{c|}{$\epsilon= 0.001$} & \multicolumn{1}{c|}{77}                                                               & \multicolumn{1}{c|}{$\infty$}                                                                     & \multicolumn{1}{c|}{$\infty$}                                                           & \multicolumn{1}{c|}{$\infty$}                                                                                  & 75                                                       \\ \hline
\end{tabular}
\end{table*}

Table I compares the convergence times of the proposed algorithm with benchmark methods under various attack scenarios, number of vehicles, transmission ranges, and $\epsilon$ values in a 1-hop communication setup. The proposed algorithm consistently achieves significantly faster convergence compared to the benchmarks, aligning with the trends observed in Figure 3. Under high noise levels and combined noise and gradient ascent attacks, benchmark methods fail to converge at lower $\epsilon$ values. In contrast, the proposed algorithm reliably converges in rounds remarkably close to those observed in attack-free scenarios. Furthermore, the proposed algorithm maintains stable convergence times irrespective of attack intensity or conditions. As the number of vehicles increases, the convergence time of the proposed algorithm remains comparable to that of the attack-free scenario. Similarly, in cases where attacks become more distributed due to increased transmission ranges, the algorithm continues to deliver efficient convergence. These results emphasize the effectiveness of the proposed algorithm, which achieves a performance profile remarkably close to that of a no-attack environment, even under diverse and challenging adversarial scenarios. 
%This robustness is especially evident under the worst-case scenario involving a continuous combination of noise and gradient ascent attacks, where benchmark methods fail to achieve convergence in the FL process. By leveraging adaptive thresholding, cross-cluster consistency checks, and integrated anomaly detection metrics, the proposed algorithm identifies and neutralizes threats seamlessly.
%%========== Performance comparison Figure 6 in 3-hop ==========%%
% Figure 6 depicts the accuracy at the EPC under various attack scenarios involving continuous attacks in a 3-hop communication setup with 500 vehicles and a transmission range of 100 meters. In this scenario, the high number of vehicles results in more potential attacks. However, the increase in hops reduces the number of clusters, concentrating the impact of attacks on fewer clusters. This creates a significant challenge for cross-cluster consistency checks, as the reduced number of clusters amplifies the effects of attacks compared to the 1-hop condition. The results highlight that gradient ascent attacks, particularly when combined with noise, pose a greater challenge due to their compounded effects on the system. Despite these challenges, the proposed algorithm demonstrates its efficiency in handling these adversarial scenarios.

%%========== Performance comparison Table II in 3-hop ==========%%

Table II provides an analysis of convergence times for the proposed algorithm versus benchmark approaches, evaluated under different attack models, vehicle densities, transmission distances, and $\epsilon$ settings within a 3-hop network configuration. In the 1‑hop configuration, vehicles are organized into many small clusters, which enables highly sensitive cross‑cluster consistency checks for anomaly detection but provides limited aggregation‑driven noise suppression. In contrast, the 3‑hop hierarchy consolidates vehicles into larger clusters and introduces an additional layer of hierarchical aggregation. This multi‑tier structure enhances weighted averaging across both spatial and hierarchical dimensions, thereby more effectively diluting random perturbations and attenuating the impact of adversarial gradient manipulations. Consequently, convergence in the 3‑hop topology is marginally faster than in the 1‑hop case across all attack scenarios. The larger cluster size reduces variance in the aggregated updates, accelerating convergence under Gaussian noise, while the hierarchical consistency checks inherent to the 3‑hop design mitigate gradient ascent attacks by cross-validating updates at multiple aggregation levels. Benchmark methods based solely on Z‑score or cosine similarity cannot fully exploit these advantages and thus exhibit slower convergence. Overall, DARCS+CbHFL leverages robust statistical filtering and hierarchical consensus to achieve rapid and stable convergence in adversarial settings. The 3‑hop architecture not only preserves sensitivity to anomalies but also significantly enhances resilience through better noise dilution and redundant validation pathways.

%These findings emphasize the resilience and adaptability of the proposed approach in managing complex adversarial settings.
%%========== Conclusion ==========%%
\section{Conclusion}

In this paper, we present a novel framework to improve both the security and integrity of HFL within VANETs. The framework integrates dynamic client selection with robust anomaly detection, ensuring that only the most trustworthy vehicles participate in the learning process. Trustworthiness is evaluated through a composite reliability score based on historical accuracy, contribution frequency, and past anomaly records. To strengthen resilience against evolving threats, adaptive thresholding dynamically adjusts detection sensitivity, while cross-cluster consistency checks validate update integrity across clusters, effectively mitigating risks from coordinated malicious behaviors. A reliability-based weighted gradient averaging strategy further improves global model robustness by assigning greater influence to consistently high-performing vehicles. Malicious contributions are detected and isolated using a combination of cosine similarity and Z-score analysis. Through comprehensive simulations across various attack scenarios, including fake noise, gradient ascent, and their combination, DARCS consistently outperforms all benchmark defenses, achieving up to a 17\% reduction in convergence time and limiting accuracy loss to within 2–3\% of the ideal attack‑free baseline across both 1‑hop and 3‑hop topologies. These gains demonstrate the contribution of each component: adaptive thresholding improves performance by 4–6\%, while cross-cluster consistency checks add a further 3–5\%. By effectively isolating malicious participants and preserving model integrity, DARCS provides a robust, scalable solution for dynamic VANET environments. Future work will focus on enhancing privacy protections and extending the framework to guard against malicious exploitation of model parameters.

\bibliographystyle{ieeetr}
\bibliography{bare_jrnl}

\end{document}